\documentclass[journal]{IEEEtran}
\usepackage{cite}
\usepackage{amsmath,graphicx}
\usepackage{algorithmic}
\usepackage[ruled,vlined]{algorithm2e}
\usepackage{changepage}
\usepackage{subcaption,siunitx,booktabs}
\usepackage{enumerate,enumitem}
\usepackage{amssymb}
\newcommand{\trans}{^{\mathsf{T}}}
\usepackage{paralist}
\usepackage{bm}
\usepackage{multirow}
\usepackage{tabularx}
\usepackage{url}
\usepackage{authblk}

\newcolumntype{L}{>{\raggedright\arraybackslash}X}
\usepackage[outdir=./]{epstopdf}

\begin{document}
\title{Generalized domain adaptation framework for parametric back-end in speaker recognition}

\author[1]{Qiongqiong Wang}
\author[2]{Koji Okabe} 
\author[1]{Kong Aik Lee}
\author[3]{Takafumi Koshinaka}
\affil[1]{Institute for Infocomm Research (I$^2$R), A$^\star$STAR, Singapore}
\affil[2]{Data Science Laboratories, NEC Corporation, Japan}
\affil[3]{School of Data Science, Yokohama City University, Japan}


%



\maketitle

\begin{abstract}
State-of-the-art speaker recognition systems comprise 
a speaker embedding front-end followed by
a probabilistic linear discriminant analysis (PLDA) back-end.
The effectiveness of these components relies on the availability of 
a large amount of labeled training data. 
In practice, it is common for domains (e.g., language, channel, demographic) in which a system is deployed to differ from that in
which a system has been trained.
To close the resulting gap, domain adaptation is often essential for PLDA models.
Among two of its variants are Heavy-tailed PLDA (HT-PLDA) and Gaussian PLDA (G-PLDA). 
Though the former better fits real feature spaces than does the latter, 
its popularity has been severely limited by its computational complexity and, especially, by the difficulty, it presents in domain adaptation, which results from its non-Gaussian property.
Various domain adaptation methods have been proposed for G-PLDA.
This paper proposes a generalized framework for domain adaptation that can be applied to both of the above variants of PLDA for speaker recognition. 
It not only includes several existing supervised and unsupervised domain adaptation
methods but also makes possible more flexible usage of
available data in different domains. 
In particular, we introduce here two new techniques: 
(1) correlation-alignment in the model level, and 
(2) covariance regularization. 
To the best of our knowledge, this is the first proposed application of such techniques for domain adaptation w.r.t. HT-PLDA. 
The efficacy of the proposed techniques has been experimentally validated on NIST 2016, 2018, and 2019 Speaker Recognition Evaluation
(SRE’16, SRE’18 and SRE'19) datasets.
\end{abstract}


%
\IEEEpeerreviewmaketitle

\section{Introduction}
%
%
%
%
\IEEEPARstart{S}{peaker} recognition is the biometric task of recognizing a person from
his/her voice on the basis of a small amount of speech utterance from that person \cite{Hansen15}. 
Speaker embeddings are fixed-length continuous-value vectors that provide succinct characterizations of speakers’ voices rendered in speech utterances. 
Recent progress in speaker recognition has achieved successful application of deep neural networks to derive deep speaker embeddings from speech utterances \cite{snyder17,variani14, snyder18, okabe18}. 
In a way similar to classical i-vectors \cite{dehak11},
deep speaker embeddings live in a relatively simple Euclidean space in which distance can be measured far more easily than with much more complex input patterns. 
Techniques such as within-class covariance normalization (WCCN) \cite{hatch06}, linear discriminant analysis (LDA) \cite{bishop06},
and probabilistic linear discriminant analysis (PLDA) \cite{prince07, ioffe06, kenny10} can also be applied.

\begin{figure}[t]
\centering
\includegraphics[width=\linewidth]{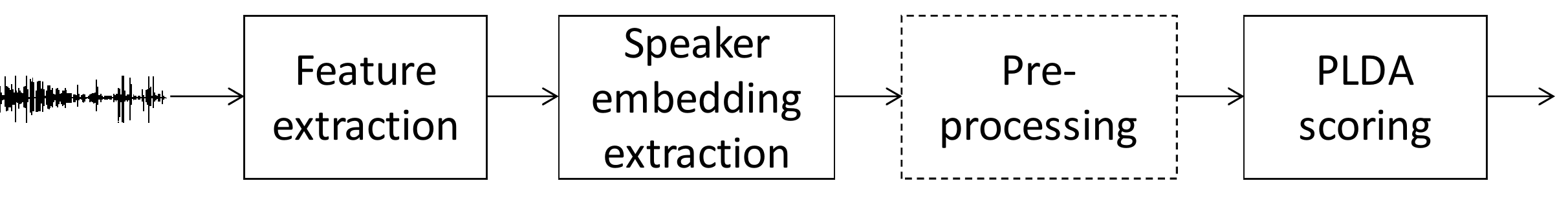}
\caption{State-of-the-art Speaker Recognition Pipeline}
\label{fig:pipeline}
\end{figure}

State-of-the-art text-independent speaker recognition systems composed of the speaker embedding front-end followed by a PLDA back-end, as illustrated in Fig.~\ref{fig:pipeline}, have shown promising performance \cite{yamamoto19}\cite{Wang22}. 
The effectiveness of these components relies on the availability of a large amount of labeled training data, 
typically over one hundred hours of speech recordings consisting of multi-session recordings from several thousand speakers
under differing conditions (e.g., w.r.t. recording devices, transmission channels, noise, reverberation, etc.).
These knowledge sources contribute to the robustness of systems against such nuisance factors. 
The challenging problem of domain mismatch arises when a speaker recognition system is used in a different domain than that of its training data (e.g., different languages, demographics, etc.).
It would be prohibitively expensive, however, to collect such a large amount of in-domain (InD) data for a new domain of interest for every application and then to retrain models. 
Most available resource-rich data that already exist will not match new domains of interest, i.e., most will be out-of-domain (OOD) data~\cite{wangqing21, Wei22, Zhang23}. 
A more viable solution would be to adapt an already trained model using a smaller, and possibly unlabeled, set of in-domain data. 

Domain adaptation techniques designed to adapt resource-rich OOD systems to produce good results in new domains have recently been studied with the aim of alleviating this problem \cite{misra14,aronowitz14,alam18,lee19b, bousquet19}.
Domain adaptation could be accomplished at different stages of the x-vector PLDA pipeline:
1) \textit{data pooling}, 
2) \textit{speaker embedding compensation}, or 
3) \textit{PLDA parameter adaptation}.
PLDA parameter adaptation is preferable in practice, as the same feature extraction and speaker embedding front-end can be used, while domain-adapted PLDA back-ends are used to cater to the condition in specific individual deployments \cite{lee20}. It directly optimizes a model in an efficient way and does not require computationally expensive retraining with large-scale OOD data or the transformation of individual feature vectors. 

There are two variants of PLDA: Gaussian PLDA (G-PLDA) and heavy-tailed PLDA (HT-PLDA). G-PLDA is the standard back-end for speaker recognizers that use speaker embeddings, assisted by a Gaussianization step involving length normalization (LN).
HT-PLDA has been shown to be superior to G-PLDA in that it better fits real embedding spaces, though the computational cost is considerable \cite{kenny10}.
It has subsequently been shown that embedding vectors could instead be Gaussianized via a simple LN procedure and provide performance comparable to that of an HT-PLDA with negligible extra computational cost \cite{garcia11}. 
Thus, G-PLDA with LN as the standard back-end for scoring text-independent speaker recognizers is the target for which model-level domain adaptation methods have been developed. 
At the same time,  HT-PLDA,  due to the computational complexity arising from its non-Gaussian property, has not achieved popularity, and it is especially difficult to adapt HT-PLDA's model parameters. 
It has, however, drawn researchers' attention because of a slightly simplified HT-PLDA with a computationally attractive
alternative to that given by G-PLDA.
This was achieved by a fast variational Bayes generative training algorithm and a fast scoring algorithm \cite{anna18}\cite{brummer18}.
The new type of HT-PLDA has demonstrated higher performance than G-PLDA + LN in NIST SRE'18 evaluations \cite{lee20}\cite{sre18}\cite{lee19a}.

It is known that PLDA models are heavily affected by domain mismatch~\cite{lee20}\cite{lee19a}\cite{matejka20}. Many domain adaptation methods have been proposed to adapt G-PLDA models independently~\cite{alam18,bousquet19,lee19b}\cite{wang20}\cite{garcia14a}.
No previous work in domain adaptation, however, has given attention to HT-PLDA.

%

Our contributions in this paper are as follows: 
First, we present consistent formulations in both embedding-level and model-level domain adaptation. Existing works have presented domain adaptation only in either embedding-level or model-level domain adaptation. In this paper, we solve the data shift problem at the model level.
Second, our comparison of embedding- and model-level domain adaptation reveals unique traits and shows that regularization is important to domain adaptation.
Third, we propose a generalized form for semi-supervised domain adaptation that offers a way to use models or embeddings, and labeled or unlabeled data.
We have also extended our previous work \cite{wang20} - correlation-alignment-based interpolation and covariance regularization on the basis of linear interpolation \cite{garcia14a} for robust domain adaptation.
Finally, we demonstrate the above-mentioned three contributions in two back-ends and enable HT-PLDA to perform domain adaptation at the model level.
We present an approximated formulation for unsupervised HT-PLDA adaptation based on state-of-the-art techniques for G-PLDA.
We also review here a standard supervised adaptation method based on linear combination~\cite{garcia14a} and show how those adaptation techniques behave in popular settings~\cite{snyder18}\cite{okabe18}\cite{yamamoto19}.
To the best of our knowledge, this is the first such work on HT-PLDA adaptation.


The remainder of this paper is organized as follows.
Section \ref{sec:relatedwork} reviews related work of domain adaptation. 
Section \ref{sec:sec2} reviews G-PLDA and HT-PLDA.
Section \ref{sec:generalized} introduces a proposed generalized framework for domain adaptation.
Section \ref{sec:sec4} compares several unsupervised and supervised domain adaptation methods and shows their relationships to the generalized framework.
Section \ref{sec:exp} describes our experimental setup, results, and analyses, and 
Section \ref{sec:summary} summarizes our work.

\section{Related Work}
\label{sec:relatedwork}

Data pooling has been proposed, for example, to add a small amount of InD data (insufficient to train a model by itself) to a large amount of OOD data to train PLDA together\cite{misra14}. Domain adaptation in this stage utilizes the speaker labels of the InD data. Thus, it is a kind of supervised domain adaptation. 
Both supervised and unsupervised domain adaptation methods can be applied to speaker embedding vectors.  In adaptations at this level, statistical information about speaker embeddings in both the OOD and InD domains is used to process the OOD embeddings in order to compensate for shifts and rotations in embedding space caused by domain mismatch.

Both supervised and unsupervised domain adaptation methods can be applied at the speaker embedding level and model level.  At the speaker embedding level, statistical information about speaker embeddings in both the OOD and InD domains is used to process the OOD embeddings in order to compensate for shifts and rotations in embedding space caused by domain mismatch.
Inter dataset variability compensation (IDVC) has been proposed to compensate for dataset shift by constraining the shifts to a low dimensional subspace\cite{aronowitz14}. 
CORrelation ALignment (CORAL) adaption whitens and re-colors OOD embeddings
using total covariance matrices calculated from OOD and InD embeddings\cite{alam18}. 
Feature-Distribution Adaptor \cite{bousquet19} and CORAL++ \cite{Li22} further apply regularization on top of the second-order statistics alignment to avoid the influence of residual components and inaccurate information during adaptation.
More recently, deep neural networks have been found to be effective in directly mapping speaker embeddings to another space \cite{Khoury22}.
At the model level, a PLDA model trained using OOD data can be adapted using either statistical information regarding 
the speaker embeddings in the InD data or parameters of another PLDA model trained using a limited amount of labeled InD data.

Among popular and effective ones are clustering methods \cite{shum14}, which label InD data using cluster labels, Kaldi domain adaptation \cite{kaldi} and CORAL+ \cite{lee19b}, which utilize the total covariance matrix calculated from unlabeled InD data to update an OOD PLDA. Among these, supervised domain adaptation is more powerful than unsupervised. In \cite{wang16}, a maximum likelihood linear transformation has been proposed for transforming OOD PLDA parameters, so as to be closer to InD. A linear interpolation method has been proposed for combining parameters of PLDAs trained separately with OOD and InD data, so as to take advantage of both PLDAs \cite{garcia14a}.
In some applications, both labeled and unlabeled InD data exist.
To make the best use of data and labels, we can apply semi-supervised domain adaptation methods. 
In \cite{wang20}, correlation-alignment-based interpolation has been proposed that utilizes the statistical information in the features of the InD data to update OOD PLDA before interpolation with an InD PLDA trained with labeled InD data. 


\section{Speaker verification in the embedding space}
\label{sec:sec2}


In state-of-the-art text-independent speaker recognition systems, the probabilistic linear discriminant analysis (PLDA) that was originally introduced in \cite{prince07}\cite{ioffe06} for face recognition has become heavily employed for speaker embedding front-end \cite{dehak11}\cite{kenny10}\cite{matejka11} (see Fig.~\ref{fig:pipeline}).
PLDA decomposes the total variability into within-speaker and between-speaker variability.
Some popular speaker embeddings are x-vectors~\cite{snyder18} that utilize the time-delayed neural network and a statistic or attentive pooling layer to obtain fixed-length utterance-level representation. More sophisticated embeddings have also been proposed such as xi-vectors~\cite{lee21} which employ a posterior inference pooling to predict the frame-wise uncertainty of the input and propagate to the embedding vector estimation. 
There are two main variants of PLDAs: Gaussian PLDA (G-PLDA) and heavy-tailed  PLDA  (HT-PLDA).

\subsection{Gaussian PLDA (G-PLDA)}
\label{sec:gplda}

G-PLDA can be seen as a special case of Joint Factor Analysis \cite{kenny08} with a single Gaussian component. 
Let $\bm{\phi}$ be a $D$-dimensional embedding vector (e.g., x-vector, xi-vector, etc.). 
We assume that vector $\bm{\phi}$ is generated from a linear Gaussian model \cite{bishop06}, as follows \cite{prince07}\cite{ioffe06}:
%
\begin{equation}
P(\bm{\phi} | \mathbf{h,x}) = \mathcal{N}(\bm{\phi} | \bm{\mu} + \bf{Fh}+ \bf{Gx}, \bf{\Sigma} ),
\label{eq:gplda}
\end{equation}
where $\bm{\mu}\in \mathbb{R}^D$ represents the global mean.
The variables $\bf{h}$ and $\bf{x}$ are, respectively, the latent speaker and channel variables of $d_h$ and $d_x$ dimensions,
while $\mathbf{F}\in \mathbb{R}^{D\times d_h}$ and $\mathbf{G}\in \mathbb{R}^{D\times d_x}$ are, respectively, the speaker and channel loading matrices, 
and the diagonal matrix $\mathbf{\Sigma}$ models residual variances. 
The between- and within-speaker covariance matrices $\{\mathbf{\Phi}_\mathrm{B}, \bf{\Phi}_\mathrm{W}\}$ can be derived from the latent speaker and channel variables as
\begin{equation}
\bf{\Phi}_\mathrm{B}  = \bf{FF} {\trans},
\bf{\Phi}_\mathrm {W} = \bf{GG}\trans +\bf { \Sigma}.
\end{equation}
The speaker conditional distributions
\begin{equation}
P(\bm{\phi}|\bf{h})=\mathcal{N}(\bm{\phi}|\bm{\mu}+\bf{Fh, \Phi}_\mathrm {w} )  \\
\end{equation}
have a common within-speaker covariance matrix $\mathbf{\Phi}_\mathrm {W}$.
The speaker variables have a Gaussian
prior:
\begin{equation}
P(\bf{h})=\mathcal{N}(\bf{h| 0, I }).  \\
\end{equation}

\subsection{Heavy-tailed PLDA (HT-PLDA)}
\label{sec:htplda}

Unlike G-PLDA, which assumes Gaussian priors of both channel and speaker factors $\mathbf{x}$ and $\mathbf{h}$ in \eqref{eq:gplda},
the heavy-tailed version of PLDA (HT-PLDA) replaces Gaussian distributions with Student’s t-distributions.
In \cite{anna18}\cite{brummer18}, a fast training and scoring algorithm for a simplified HT-PLDA back-end was proposed, 
 with speed comparable to that with G-PLDA. 
 For every speaker, a hidden speaker identity variable, $\mathbf{h}\in \mathbb{R}^{d_h}$,
 is drawn from the standard normal distribution.
The heavy-tailed behavior is obtained by having a precision scaling factor, $\lambda>0$, drawn from a gamma distribution, $ \mathcal{G}(\alpha,\beta)$ parametrized by $\alpha=\beta=\frac{\nu}{2}>0$. 
The parameter $\nu$ is known as the \textit{degrees of freedom} \cite{bishop06}\cite{kenny10}.
Given the hidden variables, 
the embedding vectors $\bm{\phi}$ are described by a multivariate normal:
\begin{equation}
P(\bm{\phi}|\mathbf{h},\lambda)
=\mathcal{N}(\bm{\phi}| \mathbf{Fh}, (\lambda\mathbf{W})^{-1})  \\
\label{eq:htplda1}
\end{equation}
where $\mathbf{F}\in \mathbb{R}^{D\times d_h}$ is the speech loading matrix, $\mathbf{W}\in \mathbb{R}^{D\times D}$ is a positive definite \textit{precision matrix}, and $D$ is the speaker embedding dimension.
The model parameters are $\nu,\mathbf{F, W}$.
This model is a simplification of the HT-PLDA model in \cite{kenny10}.
Because of the heavy-tailed speaker identity variables,
(\ref{eq:htplda1}) can be represented in a multivariate t-distribution \cite{bishop06}\cite{brummer18}:
\begin{equation}
P(\bm{\phi}|\mathbf{h})
=\mathcal{T}(\bm{\phi}|\mathbf{Fh},\mathbf{W},\nu).
\label{eq:htplda2}
\end{equation}
In the condition in which $D>d$, and $\mathbf{F{\trans}WF}$ is invertible, 
(\ref{eq:htplda2}) can be proved to be proportional to another t-distribution, with increased degrees of freedom
$\nu'=\nu+D-d$:
\begin{equation}
P(\bm{\phi}|\mathbf{h})
=\mathcal{T}(\mathbf{h|\hat{h}},\mathbf{B},\nu')
\label{eq:htplda3}
\end{equation}
where $\mathbf{\hat{h}}$ and $\mathbf{B}$ are utterance dependent:
\begin{alignat*}{2}
\mathbf{\hat{h}} &=\mathbf{B}^{-1}\mathbf{a},
& \qquad\qquad 
\mathbf{B} &=b\mathbf{B}_0,\\
\mathbf{a} &=b\mathbf{F{\trans}W\phi},
& \qquad\qquad 
b&=\frac{\nu+D-d}{\nu+\bm{\phi}\trans\mathbf{G}\bm{\phi}}, 
\end{alignat*}
while 
$\mathbf{B}_0$ and $\mathbf{G}$ are utterance-independent parameters that only need to be calculated once and are constant for all the speaker embeddings:
\begin{align}
\mathbf{B}_0 = \mathbf{F{\trans}WF}, &&
\mathbf{G} =\mathbf{W-WFB}_0^{-1}\mathbf{F{\trans}W}.
\end{align}
In this paper, we propose domain adaptation for HT-PLDA based on this algorithm.


\subsection{Domain adaptation based on speaker embeddings}
\label{ssec:ssec3}

In speaker recognition, when test data is mismatched with the domain in which the model has been trained,  performance degrades drastically. 
In practice, collecting a large amount of data with speaker labels in the target domain is expensive.
An alternative solution is to do domain adaptation from a source domain to a target domain
in which labeled training data is scarce. 
Two typical domain adaptation scenarios are:
(1) supervised domain adaptation using a small amount of in-domain (InD) data with speaker labels, and
(2) unsupervised domain adaptation using relatively larger amount of unlabeled InD data. 

The final goal of domain adaptation based on speaker embeddings is 
to perform covariance matching from a source domain to a target domain. 
In supervised domain adaptation, a new set of between-speaker and within-speaker covariance matrices for the target domain can be calculated from a limited amount of labeled data, 
though these matrices may not be very accurate or reliable. 
They are used to adapt the out-of-domain (OOD) covariance matrices \cite{garcia14a}. 
In unsupervised domain adaptation, only a total covariance matrix (i.e., the summation of the between-speaker and within-speaker covariance matrices)
can be calculated from the unlabeled data, and not the two covariance matrices separately.
The total covariance matrix is often used for adaptation \cite{lee19b}\cite{bousquet19}\cite{kaldi}\cite{sun16}. 

Domain adaptation can be applied via data, embedding compensation, and/or model adaptation.
At the data level, InD data with speaker labels is added to a pool of OOD data to train a PLDA model \cite{misra14},
which is only feasible as a supervised method.
At the embedding level, domain adaptation is used to rotate and re-scale the embeddings in the source domain by multiplying it with a transformation matrix $\mathbf{A}$ to fit a distribution better in the target domain:
\begin{equation}
\bm{\phi}^*=\mathbf{A}\bm{\phi}.
\label{eq:feat_trans}
\end{equation}
At the model level, between-speaker and within-speaker covariance matrices can be transformed or fused.
This is popular in practice    since there is no need for extra memory to store the large amount of the source domain training data other than that for the trained OOD model.  
In principle, embedding compensation here can be considered equivalent to model-level domain adaptation,
as the transformation in embedding in \eqref{eq:feat_trans} results in a change in total covariance matrices: 
\begin{equation}
\bf{\Phi}^*=A \bf {\Phi}A{\trans}.
\label{eq:ctot_trans}
\end{equation}
The equivalency stands in the condition in which the embeddings are used directly to train a back-end model without any pre-processing.
In this paper, we present speaker embedding compensation methods in the form of model adaptation
and compare them with other model level methods, utilizing the relationships shown in \eqref{eq:feat_trans} and \eqref{eq:ctot_trans}.

In both supervised and unsupervised domain adaptation, the goal is to propagate the variance, i.e., uncertainty seen in the InD data to the adapted model.
This means that regularization \cite{lee19b} between covariance matrices will be important to guarantee the uncertainty increase. 
In addition, it can also improve the robustness of some fusion-based domain adaption models against interpolation weights.
In some applications, both labeled and unlabeled InD data exist.
To make the most use of the data and the labels, we can apply semi-supervised domain adaptation methods. 
In \cite{wang20}, correlation-alignment-based interpolation was proposed that utilized the statistical information in the features of InD data to update OOD PLDA before interpolation with an InD PLDA trained with labeled InD data.

\section{Generalized Framework for PLDA adaptation}
\label{sec:generalized}

We propose a generalized framework for robust PLDA domain adaptation in both unsupervised and supervised manners:
\begin{equation}
{\bf{\Phi^+}}= \alpha {\bf{\Phi}}_{\mathrm{0}} 
+ \beta
\Gamma_\mathrm{max} ( 
{\bf{\Phi}}_{\mathrm{1}},
{\bf{\Phi}}_{\mathrm{2}}
) 
\label{eq:general}
\end{equation}
where
$\mathbf{\Phi}^+$ represents the between and within-speaker covariance matrix of the domain-adapted PLDA.
$ {\bf{\Phi}}_{\mathrm{0}} $, ${\bf{\Phi}}_{\mathrm{1}}$ and ${\bf{\Phi}}_{\mathrm{2}}$ 
are covariance matrices from up to three PLDA models, and $\{\alpha, \beta\}$ are the weighting parameters.
The regularization function 
\begin{equation}
\Gamma_\mathrm{max}({\bf {\bf{\Phi}}_{\mathrm{1}}}, {\bf {\bf{\Phi}}_{\mathrm{2}}})=
{\mathbf{B}} {\trans} 
\max(\bf{ E, I}) {\bf{ B}} 
\label{eq:max} \\
\end{equation}
can be considered a processed covariance matrix computed taking the information of both covariance matrices and that guarantees the variability to be larger than both ${\bf{\Phi}}_{\mathrm{1}}$ and ${\bf{\Phi}}_{\mathrm{2}}$.
The operator $\max(\cdot)$ takes the larger values from two matrices in an element-wise manner. 
The diagonal matrix $\bf{E}$ and the identical matrix $\bf{I}$ correspond to $\{\bf{\Phi}_1,\bf{\Phi}_2\}$, respectively, after the projection of $\mathbf{B}\trans(\cdot)\mathbf{B}$:
\begin{align}
{\bf{B}}{\trans} \bf {\bf{\Phi}}_{\mathrm{1}} \bf{B} = E &&
{\bf{B}}{\trans} \bf {\bf{\Phi}}_{\mathrm{2}} \bf{B} = I.
\end{align}
In the case where the between- and within-speaker covariance matrices are real-value symmetric matrices with full ranks, 
 they can be eigen-decomposed (EVD) with $d$ number of real non-zero eigenvalues and corresponding eigenvectors. Thus,  $\bf{\{B,E\}}$ are obtained by a simultaneous diagonalization of $\{\bf{\Phi}_1,\bf{\Phi}_2\}$~\cite{horn13}: 
 \begin{align*}
\{\mathbf{Q,\Lambda\}} &\leftarrow \mathrm{EVD} (\mathbf{\Phi}_1) \\
\{\mathbf{P,E\}} &\leftarrow \mathrm{EVD} (\mathbf{\Lambda} ^ \mathrm{-1/2} \mathbf{Q}{\trans} \mathbf{\Phi}_2 \mathbf{Q} \Lambda ^ \mathrm{-1/2}) \\
\bf{B}&=\mathbf{Q\Lambda} ^ \mathrm{-1/2} \mathbf{P}
\end{align*}
$\bf{B}$ is shown to be a square matrix with real-value elements and $\bf{BB}\trans=I$, and thus, is an orthogonal matrix. 

The generalized framework \eqref{eq:general} is a linear interpolation of ${\mathbf{\Phi}}_0$ and a regularization result from ${\mathbf{\Phi}}_1$ and ${\mathbf{\Phi}}_2$.
To have a better interpretation, we define ${\mathbf{\Phi}}_0$ as a covariance matrix of a base PLDA from which a new PLDA has been adapted. Though ${\bf{\Phi}}_{\mathrm{1}}$ and ${\bf{\Phi}}_{\mathrm{2}}$, theoretically, can be swapped, we interpret them differently. 
When a new PLDA (corresponding to $\mathbf{\Phi}_1$ here) is available to develop the base PLDA (${\mathbf{\Phi}}_0$), instead of a direct linear interpolation between them, we regularize $\mathbf{\Phi}_1$ using the covariance $\mathbf{\Phi}_2$ of another PLDA as the reference. Thus, PLDAs that correspond to $\mathbf{\Phi}_1$ and $\mathbf{\Phi}_2$ are defined as a developer PLDA and a reference PLDA, respectively. 
This generalized framework enables us to combine several existing supervised  and  unsupervised  domain  adaptations into  a  single  formulation,
which we show in Section~\ref{sec:sec4} that follows.


For G-PLDA where the dimension is set to the same as that of the embedding vectors, with the number of speakers in the training data larger than the dimension, the between- and within-speaker covariance matrices are real-value symmetric matrices with full ranks. Therefore, regularization (\ref{eq:max}) can be applied as mentioned above.
Next, let us give an estimation of domain adaptation for HT-PLDA.
In the new algorithm of HT-PLDA  introduced in Section~\ref{sec:htplda}, 
the approximate between- and within-speaker covariance matrices are
\begin{align}
{\bf{\Phi_B}} ={\bf{F}} {\bf{F}}{\trans}, &&
{\bf{\Phi_W}}
=(\lambda {\bf{W}})^{-1}.
\label{eq:ht_bw}
\end{align}
It should be noted that the within-speaker covariance matrix is a variable dependent on individual utterances because 
$\lambda$ is the hidden precision scaling factor that varies utterance by utterance.
We assume that the scalar parameter $\nu$ and matrix $\bf{W}^{-1}$ are independent of each other, and we approximate the domain adaptation of the within-speaker covariance matrix as the adaptation of the matrix $\bf{W}^{-1}$ and the scalar parameter $\nu$ separately. 
We propose to adapt $\bf{W}^{-1}$ in the same way as with the between- and within-speaker covariance matrices in \eqref{eq:general}.
In this paper, we do not discuss the adaptation of the parameter $\nu$, as \cite{anna18}\cite{brummer18} have shown it to be a pre-defined parameter. 

Unlike the within-speaker covariance matrix, the between-speaker covariance in (\ref{eq:ht_bw}) is a fixed matrix once the HT-PLDA model has been trained. 
Because the new algorithm for HT-PLDA \cite{anna18}\cite{brummer18} is based on the assumption of $d_h \ll D$, the between-speaker covariance matrix is rank-deficient. 
This causes the inapplicability of the proposed covariance regularization technique in \eqref{eq:max}.
Studies have shown that between-speaker covariance is less affected by domain mismatch than is within-speaker covariance.
Thus, in this paper for HT-PLDA, we investigate the effect of the proposed covariance regularization technique only in the within-speaker covariance matrix.


\section{Relationships and extensions to existing methods}
\label{sec:sec4}
\begin{figure}[t]
\centering
\includegraphics[width=\linewidth]{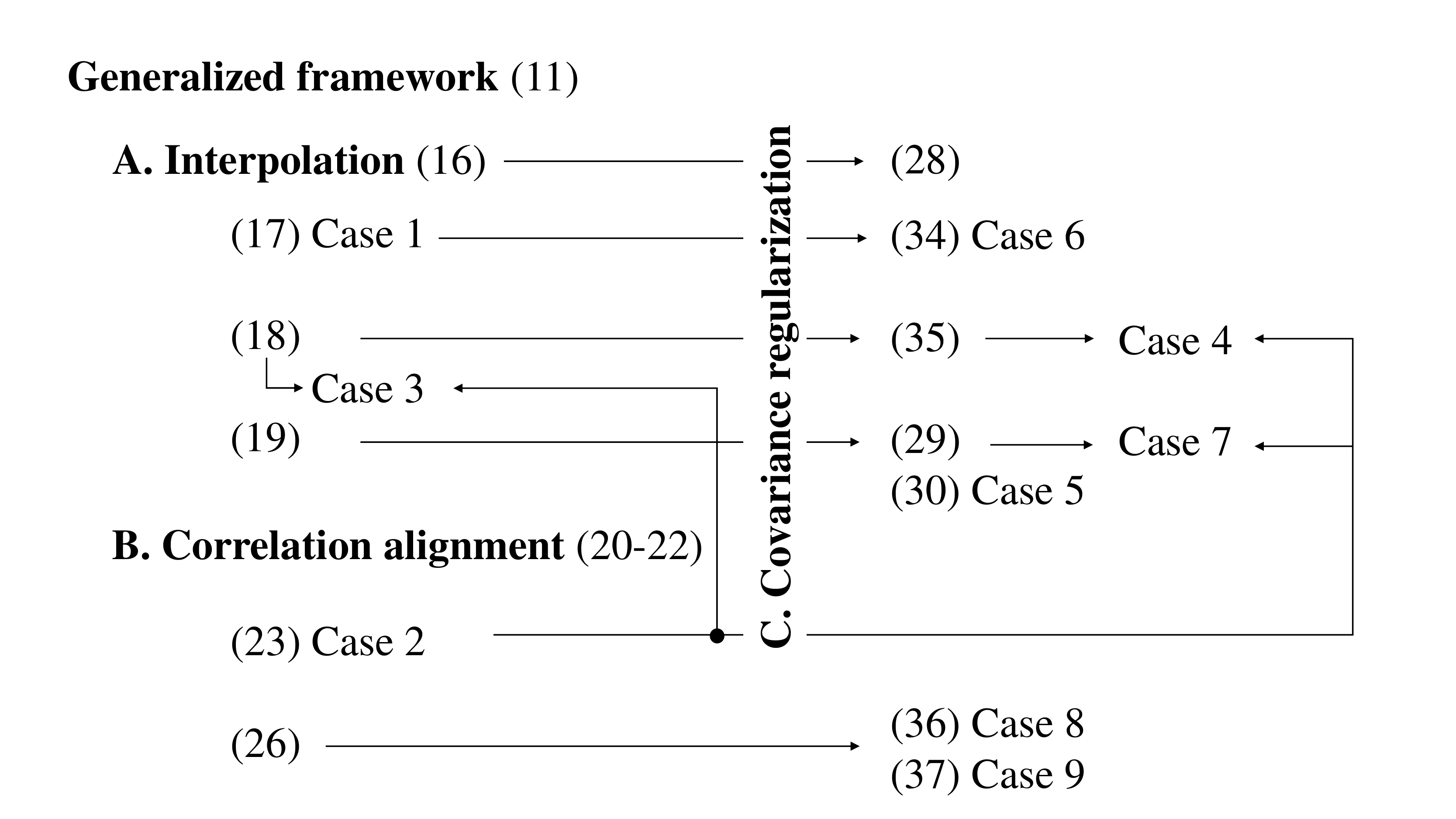}
\caption{The relationships between the special cases from the perspective of the three main factors of the generalized framework. (*) shows the equation index.}
\label{fig:relation}
\end{figure}

The generalized framework  can be summarized in terms of the next three main factors:
\begin{inparaenum}[i)]
    \item interpolation of covariance matrices, 
    \item correlation alignment, and
    \item covariance regularization.
\end{inparaenum}
In this section, we introduce several existing special cases of the general framework, as listed in Tab.~\ref{tab:generalform1}, and show their relationships with each other from the perspective of the three factors. The relationship diagram is briefly shown in Fig.~\ref{fig:relation}, and their corresponding equations mentioned in this Section are listed in Tab.~\ref{tab:eq}) for fast reference.

\begin{table}[t] 
\centering
\bgroup
\def\arraystretch{1.1}%
\setlength\tabcolsep{2.0pt}
\caption{The special cases derived from the general form.}
\begin{tabular}{c|*{4}{l}r}
\hline
& Method & $\Phi_{0}$ & $\Phi_{1}$ & $\Phi_{2}$ & Eq. \\
\hline \hline
1 & LIP \cite{garcia14a} & ${\bf{\Phi}}_{\mathrm{I}}$ & ${\bf{\Phi}}_{\mathrm{O}}$ & ${\bf{\Phi}}_{\mathrm{O}}$ & \eqref{eq:lip} \\
2 & CORAL \cite{alam18}&  0 & ${\bf{\Phi}}_{\mathrm{CORAL}}$& ${\bf{\Phi}}_{\mathrm{CORAL}}$& \eqref{eq:coral_feat}\\
3 & CIP  \cite{wang20}& ${\bf{\Phi}}_{\mathrm{I}}$ & ${\bf{\Phi}}_{\mathrm{CORAL}}$ & ${\bf{\Phi}}_{\mathrm{CORAL}}$ & \eqref{eq:lip_pseudo1}\eqref{eq:coral_align3} \\
4 & CORAL+\cite{lee19b} & ${\bf{\Phi}}_{\mathrm{O}}$ & ${\bf{\Phi}}_{\mathrm{CORAL}}$ & ${\bf{\Phi}}_{\mathrm{O}}$ &\eqref{eq:lip_reg_O}\eqref{eq:coral_align3}\\
5 & Kaldi\cite{kaldi} & ${\bf{\Phi}}_{\mathrm{O}}$ &$ {\bf{C}}_\mathrm{I}$ & ${\bf{\Phi}}_{\mathrm{O}}^b + {\bf{\Phi}}_{\mathrm{O}}^w $ & \eqref{eq:kaldi} \\
6 & LIP reg \cite{wang20} & ${\bf{\Phi}}_{\mathrm{I}}$ & ${\bf{\Phi}}_{\mathrm{O}}$ & ${\bf{\Phi}}_{\mathrm{I}}$ & \eqref{eq:lip_reg_IO} \\
7 & CIP reg  \cite{wang20}& ${\bf{\Phi}}_{\mathrm{I}}$ & ${\bf{\Phi}}_{\mathrm{CORAL}}$ & ${\bf{\Phi}}_{\mathrm{I}}$ & \eqref{eq:lip_reg_I}\eqref{eq:coral_align3} \\


8 & FDA \cite{bousquet19} &  0 & ${\bf{\Phi}}_{\mathrm{I,pseudo,1}}$& ${\bf{\Phi}}_{\mathrm{I,pseudo,1}}$& \eqref{eq:fda}\\
9 & Kaldi*~\cite{bousquet19}&  0 & ${\bf{\Phi}}_{\mathrm{I,pseudo,2}}$& ${\bf{\Phi}}_{\mathrm{I,pseudo,2}}$& \eqref{eq:coral_align3}\eqref{eq:mk2}\\

\hline
\end{tabular}
\label{tab:generalform1}
\egroup
\end{table}
\subsection{Interpolation of covariance matrices}
\label{ssec:interpolation}
%
In a special case in which ${\bf{\Phi}}_{\mathrm{1}} = {\bf{\Phi}}_{\mathrm{2}}$ in \eqref{eq:general}, meaning that they are from the same PLDA model, the regularization term $\Gamma_{max} (\cdot)$ is equivalent to the covariance matrix itself. Thus, the generalized framework presents a simple interpolation with a weight $\alpha$ of PLDA parameters, i.e., between- and within-speaker covariance of two  PLDAs
\begin{equation}
{\bf{\Phi^+}}= \alpha {\bf{\Phi}}_{\mathrm{0}} 
+ \beta{\bf{\Phi}}_{\mathrm{1}}.
\label{eq:general1}
\end{equation}
%

In common domain adaptation scenarios, a sufficient amount of labeled OOD data are always available. InD data with labels is often insufficient, while unlabeled InD data is easier to access and obtain.
In the first scenario, in which the InD data is labeled but insufficient, 
an InD PLDA model can be trained but will be rather unreliable or not be expected to perform well. It can, however, be interpolated with an OOD PLDA model   
\begin{equation}
\bf{\Phi^+}= \alpha {\bf{\Phi}}_{\mathrm{I}} 
+ (1-\alpha){\bf{\Phi}}_{\mathrm{O}}
\label{eq:lip}
\end{equation}
and play an important role leading the OOD model in the direction of the target domain. It has been proposed as a supervised method in \cite{garcia14a} and is shown as special case 1 linear interpolation (LIP) in Tab.~\ref{tab:generalform1}. 

For the same scenario, the interpolation can be also applied to the InD model with a  pseudo-InD model 
\begin{equation}
\bf{\Phi^+}= \alpha {\bf{\Phi}}_{\mathrm{I}} 
+ (1-\alpha){\bf{\Phi}}_{\mathrm{I,pseudo}},
\label{eq:lip_pseudo1}
\end{equation}
where $\bf{\Phi}_{\mathrm{I,pseudo}}$ represents the preliminarily adapted PLDA model.
It is supposed to be closer to a true InD PLDA and fit the target domain better than an OOD PLDA. Therefore, replacing the OOD model in \eqref{eq:lip} with the pseudo-Ind model is expected to result in a more reliable PLDA. 
More about the pseudo-InD model will be introduced later in this section.

In the second scenario, in which the InD data is unlabeled, an interpolation still can be conducted between an OOD PLDA and a pseudo-IND PLDA
\begin{equation}
\bf{\Phi^+}= \alpha {\bf{\Phi}}_{\mathrm{O}}  
+ (1-\alpha){\bf{\Phi}}_{\mathrm{I,pseudo}}.
\label{eq:lip_pseudo2}
\end{equation}
The pseudo-InD PLDA here is a preliminarily adapted model from an OOD PLDA model using InD data in an unsupervised manner.


\begin{table}[t] 
\centering
\bgroup
\def\arraystretch{1.5}%
\setlength\tabcolsep{3.0pt}
\caption{Relevant equation for fast reference for the relationship diagram in Fig. \ref{fig:relation}.}
\begin{tabular}{l|c|l} 
\hline
 & Eq. & Equation   \\
\hline \hline
 \multirow{2}{*}{\begin{minipage}{0.7in}Generalized framework \end{minipage}} 
 & \multirow{2}{*}{\ref{eq:general}} & \multirow{2}{*}{${\bf{\Phi^+}}= \alpha {\bf{\Phi}}_{\mathrm{0}} 
+ \beta\Gamma_\mathrm{max} \left(
{\bf{\Phi}}_{\mathrm{1}},
{\bf{\Phi}}_{\mathrm{2}},
\right) $} \\
& & \\
\hline
\multirow{4}{*}{A. Interpolation}  & \ref{eq:lip} & $\bf{\Phi^+}= \alpha {\bf{\Phi}}_{\mathrm{I}} 
+ (1-\alpha){\bf{\Phi}}_{\mathrm{O}}  $\\
\cline{2-3}
 & \ref{eq:lip_pseudo1}& $\bf{\Phi^+}= \alpha {\bf{\Phi}}_{\mathrm{I}} 
+ (1-\alpha){\bf{\Phi}}_{\mathrm{I,pseudo}} $ \\
\cline{2-3}
 & \ref{eq:lip_pseudo2} & $\bf{\Phi^+}= \alpha {\bf{\Phi}}_{\mathrm{O}}  
+ (1-\alpha){\bf{\Phi}}_{\mathrm{I,pseudo}}$  \\
\cline{2-3}
 & \ref{eq:coral_align2} & $\bm{\phi}_\mathrm{I, pseudo} = \mathbf{A} \bm{\phi}_\mathrm{O} $ \\
\hline 
\multirow{4}{*}{\begin{minipage}{0.7in}B. Correlation alignment\end{minipage}}  & \ref{eq:coral_align1} &$\bf{C}_{\mathrm{I,pseudo}} = \mathbf{A}  \bf{C}_{\mathrm{O}} \mathbf{A}{\trans} = \bf{C}_{\mathrm{I}} $ \\
\cline{2-3}
& \ref{eq:coral_align_apa}& $\bf{\Phi}_{\mathrm{I,pseudo}} = \mathbf{A} {\bf{\Phi}}_{\mathrm{O}}\mathbf{A}{\trans} $\\
\cline{2-3}
& \ref{eq:coral1}& $\mathbf{A} = \mathrm{C}_\mathrm{I}^\frac{1}{2} \mathbf{C}_\mathrm{O}^{-\frac{1}{2}}$\\
\cline{2-3}
& \ref{eq:coral2}& $\mathbf{C}_\mathrm{O}^{-\frac{1}{2}}
=\mathbf{Q}_\mathrm{O} \mathbf{\Lambda}_\mathrm{O}^{-\frac{1}{2}}
\mathbf{Q}_\mathrm{O}{\trans}, 
\mathbf{C}_\mathrm{I}^{\frac{1}{2}}
=\mathbf{Q}_\mathrm{I}\mathbf{\Lambda}_\mathrm{I}^{\frac{1}{2}}
\mathbf{Q}_\mathrm{I}{\trans}$\\
\cline{2-3}
& \ref{eq:fda1}& $\bm{\phi}_\mathrm{I, pseudo} = 
\mathrm{\mathbf{C}_O^{\frac{1}{2}}
\mathbf{P}{\Delta}^{\frac{1}{2}} \mathbf{P}^T 
\mathbf{C}_O ^{-\frac{1}{2}}}
\bm{\phi}_\mathrm{O}$ \\
\hline
\multirow{7}{*}{\begin{minipage}{0.7in}C. Covariance regularization\end{minipage}} &\ref{eq:reg1} & ${\bf{\Phi^+}}= \alpha {\bf{\Phi}}_{\mathrm{0}} 
+ (1-\alpha)
\mathbf{\Gamma}_\mathrm{max} ( 
{\bf{\Phi}}_{\mathrm{1}},
{\bf{\Phi}}_{\mathrm{0}}
) $\\
\cline{2-3}
& \ref{eq:lip_reg_O} & $\bf{\Phi^+}= \alpha {\bf{\Phi}}_{\mathrm{O}}  
+ (1-\alpha)\Gamma_\mathrm{max} ( {\bf{\Phi}}_{\mathrm{I,pseudo}},{\bf{\Phi}}_{\mathrm{O}})$\\
\cline{2-3}
& \ref{eq:kaldi}& $\bf{\Phi}_\mathrm {tot} ^{+} = \alpha \bf {\Phi}_\mathrm {tot,O} 
+ (1-\alpha) \Gamma_\mathrm{max} ( 
{\bf{C}}_\mathrm {I},
{\bf{\Phi}}_\mathrm {tot,O}
) $\\
\cline{2-3}
& \ref{eq:lip_reg_IO}& $\bf{\Phi^+}= \alpha {\bf{\Phi}}_{\mathrm{I}}  
+ (1-\alpha)\Gamma_\mathrm{max} ( {\bf{\Phi}}_{\mathrm{O}},{\bf{\Phi}}_{\mathrm{I}}) $\\
\cline{2-3}
&\ref{eq:lip_reg_I}& $\bf{\Phi^+}= \alpha {\bf{\Phi}}_{\mathrm{I}}  
+ (1-\alpha)\Gamma_\mathrm{max} ( {\bf{\Phi}}_{\mathrm{I,pseudo}},{\bf{\Phi}}_{\mathrm{I}} ) $\\
\cline{2-3}
& \ref{eq:fda} & $\bm{\phi}_\mathrm{I, pseudo}  = 
\mathrm{\mathbf{C}_O^{\frac{1}{2}}
\mathbf{P}\hat{\Delta}^{\frac{1}{2}} \mathbf{P}{\trans} 
\mathbf{C}_O ^{-\frac{1}{2}}}
\bm{\phi}_\mathrm{O}$\\
\cline{2-3}
& \multirow{2}{*}{\ref{eq:mk2}}  & $\bm{\phi}_\mathrm{I, pseudo}  = \mathbf{\Phi}_\mathrm{tot,O}^{\frac{1}{2}}\mathbf{P}\hat{\Delta}^{\frac{1}{2}} \mathbf{P}{\trans} \mathbf{\Phi}_\mathrm{tot,O} ^{-\frac{1}{2}}\bm{\phi}_\mathrm{O}$ \\
& & $\mathbf{P}\hat{\Delta}\mathbf{P}{\trans}=\Gamma_{max} ( \mathbf{\Phi}_\mathrm{tot,O} ^{-\frac{1}{2}} \mathbf{C}_\mathrm{I}  \mathbf{\Phi}_\mathrm{tot,O} ^{-\frac{1}{2}}, \mathbf{I})$\\
\hline
\end{tabular}

\label{tab:eq}
\egroup
\end{table}
\subsection{Correlation alignment}
\label{ssec:correlation}
Covariance alignment aims to align second-order statistics, i.e., covariance matrices, of OOD embeddings 
so that they will match InD embeddings 
while maintaining the good properties that OOD PLDA learned from a large amount of data. 
In formulaic terms, it would be to find a matrix $\bf{A}$ so that the pseudo-InD embeddings $\bm{\phi}_\mathrm{I, pseudo}$ that are transformed from the OOD embeddings 
\begin{equation}
\bm{\phi}_\mathrm{I, pseudo} = \mathbf{A} \bm{\phi}_\mathrm{O}
\label{eq:coral_align2}
\end{equation}
will have the total covariance matrix $\mathbf{A}  \bf{C}_{\mathrm{O}} \mathbf{A}{\trans}$ close to or almost equivalent to that of InD. To calculate the transformation matrix $\mathbf{A}$, we set 
\begin{equation}
\bf{C}_{\mathrm{I,pseudo}} \leftarrow \mathbf{A}  \bf{C}_{\mathrm{O}} \mathbf{A}{\trans} = \bf{C}_{\mathrm{I}},
\label{eq:coral_align1}
\end{equation}
where $ \bf{C}_{\mathrm{O}}$ and  $ \bf{C}_{\mathrm{I}}$ are the empirical total covariance matrices estimated from the OOD dataset $X_\mathrm{O}$ and InD data sets $X_\mathrm{I}$.
It is commonly known that a linear transformation on a normally distributed vector leads to an equivalent transformation on the mean vector and covariance matrix of its density function. 
Trained with the pseudo-InD embeddings obtained in \eqref{eq:coral_align2}, the between- and within-speaker covariance matrices of the pseudo-InD PLDA have the corresponding transformation  
\begin{equation}
\bf{\Phi}_{\mathrm{I,pseudo}} = \mathbf{A} {\bf{\Phi}}_{\mathrm{O}}\mathbf{A}{\trans}.
\label{eq:coral_align_apa}
\end{equation}
Therefore, in this paper, we consider any embedding-level adaptation using a transformation to be equivalent to model-level adaptation.


To satisfy \eqref{eq:coral_align1}, one ample algorithm is
\begin{equation}
\mathbf{A} = \mathrm{C}_\mathrm{I}^\frac{1}{2} \mathbf{C}_\mathrm{O}^{-\frac{1}{2}},
\label{eq:coral1}
\end{equation}
%
which consists of whitening $\mathbf{C}_\mathrm{O}^{-\frac{1}{2}}$ followed by re-coloring $\mathbf{C}_\mathrm{I}^{\frac{1}{2}}$:
\begin{align}
\mathbf{C}_\mathrm{O}^{-\frac{1}{2}}
=\mathbf{Q}_\mathrm{O} \mathbf{\Lambda}_\mathrm{O}^{-\frac{1}{2}}
\mathbf{Q}_\mathrm{O}{\trans}, &&
\mathbf{C}_\mathrm{I}^{\frac{1}{2}}
=\mathbf{Q}_\mathrm{I}\mathbf{\Lambda}_\mathrm{I}^{\frac{1}{2}}
\mathbf{Q}_\mathrm{I}{\trans}.
\label{eq:coral2}
\end{align}
Here, $\mathbf{Q}$ and $\mathbf{\Lambda}$ are the eigenvectors
and eigenvalues pertaining to the covariance matrices. 

\begin{figure}[t]
\centering
\includegraphics[width=\linewidth]{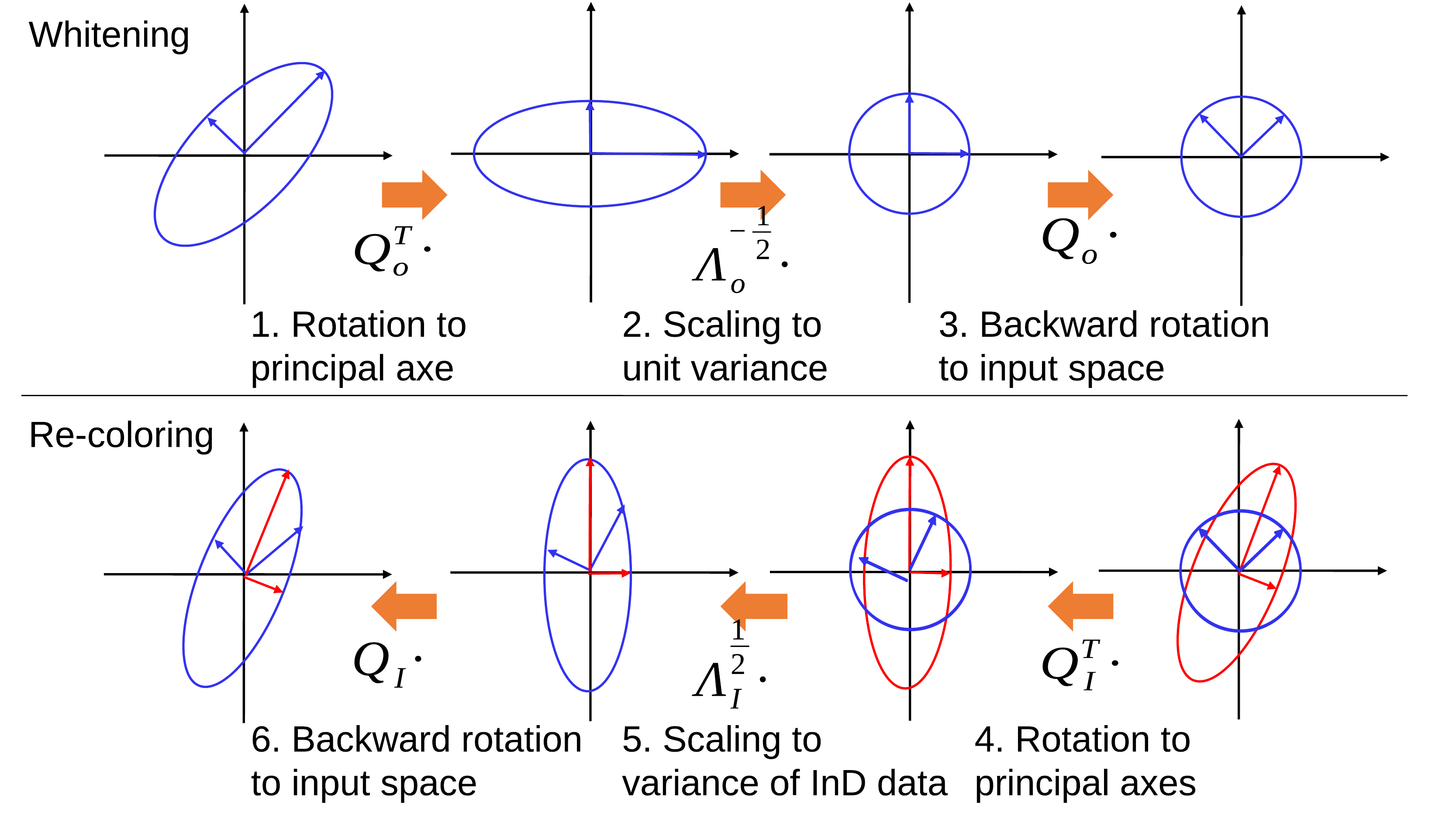}
\caption{Illustration of whitening and re-coloring with ZCA transformations in CORAL }
\label{fig:zca}
\end{figure}
Thus the whitening and re-coloring processes can be interpreted as the zero-phase
component analysis (ZCA) transformation \cite{Kessy18} with six steps, as shown in Fig.~\ref{fig:zca}.
For the whitening process, the OOD embeddings $\bm{\phi}_\mathrm{O}$ are first rotated to the principal axes of $\mathbf{C}_\mathrm{O}$ space, then scaled into a unit variance, and finally rotated back to the input space.
For the re-coloring process, similarly, the normalized embeddings are first rotated to the principal axes of $\mathbf{C}_\mathrm{I}$ space,
then re-scaled to have the variance of InD data,
and, finally, rotated backward to the input space.
As opposed to principal component analysis (PCA) and Cholesky whitening (and re-coloring), ZCA preserves the maximal similarity of the transformed feature to the original space.
Since no speaker label is used, the embedding-level domain adaptation 
\begin{equation}
\bm{\phi}_\mathrm{I, pseudo} =\mathrm{C}_\mathrm{I}^\frac{1}{2} \mathbf{C}_\mathrm{O}^{-\frac{1}{2}} \bm{\phi}_\mathrm{O}
\label{eq:coral_feat}
\end{equation}
using \eqref{eq:coral1} in \eqref{eq:coral_align2}
is an unsupervised embedding-level domain adaptation, and it is referred to as correlation alignment (CORAL) \cite{sun16}. It is shown as special case 2 in Tab.~\ref{tab:generalform1}.
Thus, when the PLDA trained using CORAL adapted embeddings as a pseudo-InD model
\begin{equation}
\bf{\Phi}_\mathrm{I,pseudo} = \mathrm{C}_\mathrm{I}^\frac{1}{2} \mathbf{C}_\mathrm{O}^{-\frac{1}{2}} 
\bf{\Phi}_\mathrm{O} 
\mathbf{C}_\mathrm{O}^{-\frac{1}{2}}\mathrm{C}_\mathrm{I}^\frac{1}{2}
\label{eq:coral_align3}
\end{equation}
for the linear interpolation with an InD model in \eqref{eq:lip_pseudo1},
the process is referred to as correlation-alignment-based interpolation (CIP), as is proposed in \cite{wang20}.  It is shown as special case 3 in Tab.~\ref{tab:generalform1}.

Another embedding transformation which still satisfies the correlation alignment in \eqref{eq:coral_align1} is
\begin{equation}
\bm{\phi}_\mathrm{I, pseudo} = 
\mathrm{\mathbf{C}_O^{\frac{1}{2}}
\mathbf{P}{\Delta}^{\frac{1}{2}} \mathbf{P}^T 
\mathbf{C}_O ^{-\frac{1}{2}}}
\bm{\phi}_\mathrm{O},
\label{eq:fda1}
\end{equation}
where $\mathbf{P}$ and $\Delta$ are from an eigenvalue decomposition 
\begin{equation}
\mathrm{ \mathbf{P}\Delta \mathbf{P}^T = \mathbf{C}_O^{-\frac{1}{2}}\mathbf{C}_I \mathbf{C}_O^{-\frac{1}{2}}}.
\label{eq:svd}
\end{equation}
In the transformation, it also uses the empirical total covariance matrices from the OOD data and InD data. 
It first whitens the features with $\mathbf{C}_\mathrm{O}^{-\frac{1}{2}}$, just as that in CORAL, 
and then recolors them with $\mathbf{C}_\mathrm{O}^{\frac{1}{2}}[\mathbf{C}_\mathrm{O}^{-\frac{1}{2}}\mathbf{C}_\mathrm{I} \mathbf{C}_\mathrm{O}^{-\frac{1}{2}}]^\frac{1}{2}$.
This is the fundamental idea of the feature-distribution adaptor (FDA) proposed in \cite{bousquet19}.


\subsection{Covariance regularization}

The central idea in domain adaptation is to propagate the uncertainty seen in the data to the model. 
Neither the adaptation equations for linear interpolation in Section~\ref{ssec:interpolation} nor those for correlation alignment in Section~\ref{ssec:correlation} guarantee that the variance, and therefore the uncertainty, will increase.  The domain adaptations for PLDA work with multiple covariance matrices, including the total covariance calculated from the available data sets and the between- and within-speaker covariance of the trainable models. Thus, regularization is introduced to propagate the uncertainty seen in any two PLDA models. As shown in Section~\ref{sec:generalized}, $\Gamma_{max}({\bf {\bf{\Phi}}_{\mathrm{1}}}, {\bf {\bf{\Phi}}_{\mathrm{2}}})$ is a regularization function that guarantees the variability to be larger than both
${\bf{\Phi}}_{\mathrm{1}}$ and ${\bf{\Phi}}_{\mathrm{2}}$. 
This insures the lower boundary of the performance of domain adaptation methods.  

\subsubsection{Covariance regularization for linear interpolations}
\label{ssseto beref4lip}
For linear interpolation-based methods, performance is strongly affected by interpolation weights, which should be correlated to the performance of the individual model. In practice, however, tuning weights is often unfeasible. 
Applying covariance regularization to the linear interpolation in \eqref{eq:general1}
\begin{equation}
{\bf{\Phi^+}}= \alpha {\bf{\Phi}}_{\mathrm{0}} 
+ (1-\alpha)
\mathbf{\Gamma}_\mathrm{max} ( 
{\bf{\Phi}}_{\mathrm{1}},
{\bf{\Phi}}_{\mathrm{0}}
) 
\label{eq:reg1}
\end{equation}
guarantees the adapted model to have a variance larger than the reference model  ${\bf{\Phi}}_{\mathrm{0}}$. 
In the case in which one covariance is larger than the other in their common space for all dimensions, the two extreme cases for \eqref{eq:reg1} are 
1) a simple interpolation ${\bf{\Phi^+}}= \alpha {\bf{\Phi}}_{\mathrm{0}} + (1-\alpha) {\bf{\Phi}}_{\mathrm{1}}$, and
2) no domain adaptation: ${\bf{\Phi^+}}= \alpha {\bf{\Phi}}_{\mathrm{0}} + (1-\alpha) {\bf{\Phi}}_{\mathrm{0}}={\bf{\Phi}}_{\mathrm{0}}$. 

Regularization is expected to take a more important role when the interpolation is done in an unsupervised manner, as, for example, between an OOD PLDA model and an unsupervised pseudo-InD PLDA in \eqref{eq:lip_pseudo2}
\begin{equation}
\bf{\Phi^+}= \alpha {\bf{\Phi}}_{\mathrm{O}}  
+ (1-\alpha)\Gamma_\mathrm{max} ( {\bf{\Phi}}_{\mathrm{I,pseudo}},{\bf{\Phi}}_{\mathrm{O}}   ).
\label{eq:lip_reg_O}
\end{equation}
When CORAL~\cite{sun16} is used to train the pseudo-InD PLDA as shown in \eqref{eq:coral_align3} for the pseudo-InD between- and within- speaker covariance matrices ${\bf{\Phi}}_{\mathrm{I,pseudo}}$, \eqref{eq:lip_reg_O} is referred as CORAL+ proposed in \cite{lee19b}.  It is shown as special case 4 in Tab.~\ref{tab:generalform1}.  The procedure is illustrated in 
Algorithm~\ref{alg:coralp} in Appendix A.


\begin{algorithm}[t]
\SetAlgoLined
  \textbf{Input} 
    Out-of-domain PLDA matrices $\{\bf{\Phi}_\mathrm{B,O},\bf{\Phi}_\mathrm{W,O}\}$\\
    \hspace*{\algorithmicindent} In-domain data $X_{I}$\\
    \hspace*{\algorithmicindent}Adaptation hyper-parameters $\{\gamma,\beta\}$ \\
  \textbf{Output} 
    Adapted covariance matrices $\{\bf{\Phi}_\mathrm{B},\bf{\Phi}_\mathrm{W}\}$\\

Estimate empirical covariance matrix from the in-domain data $X_{I}$\\
\hspace*{\algorithmicindent}    $\mathbf{C}_\mathrm{I}=\textsc{Cov}$$(X_{I})$\\
Compute out-of-domain covariance matrix\\
\hspace*{\algorithmicindent}   $\bf {\Phi}_\mathrm {tot,O}=\bf {\Phi}_\mathrm {B,O} +\bf{\Phi}_\mathrm {W,O}$ \\
Find $\bf{\{B,E\}}$ via simultaneous diagonalization of $\bf{\Phi}_\mathrm{tot,O}$ and $\bf{C}_\mathrm{I}$ \\
\hspace*{\algorithmicindent}$\{\bf{Q,\Lambda\}} \leftarrow EVD (\bf{C}_\mathrm{O})$ \\
\hspace*{\algorithmicindent}$\{\bf{P,E\}} \leftarrow EVD (\bf{\Lambda} ^ \mathrm{-1/2} Q{\trans} C_\mathrm{I} Q \Lambda ^ \mathrm{-1/2})$ \\
\hspace*{\algorithmicindent}$\bf{B=Q\Lambda} ^ \mathrm{-1/2} P$ \\
Kaldi unsupervised adaptation of PLDA, $\gamma+\beta <=1$ \\
\hspace*{\algorithmicindent}$\bf{\Phi}_\mathrm{W}^{+} = \bf{\Phi}_\mathrm{W,O} + \beta_\mathrm{W} \bf{B}^{\mathsf{-T}}\max\bf{(E-I)B}^\mathrm{-1} $\\
\hspace*{\algorithmicindent}$\bf{\Phi}_\mathrm{B}^{+} = \bf{\Phi}_\mathrm{B,O} + \beta_\mathrm{B} \bf{B}^{\mathsf{-T}} \max \bf{(E-I) B}^\mathrm{-1}.$\\
\textbf{Notation} $\mathrm{EVD}(\cdot)$ returns a matrix of eigenvectors and the corresponding eigenvalues in a diagonal matrix.
 \caption{The Kaldi algorithm \cite{kaldi} for unsupervised adaptation of PLDA.}
\label{alg:kaldi}
\end{algorithm}

Rather than predicting the between- and within-speaker covariance directly but independently, we can alternatively first predict the total covariance $\mathrm {\Phi_{tot}}$ of the adapted model
\begin{equation}
\begin{split}
\bf{\Phi}_\mathrm {tot} ^{+}& = \alpha \bf {\Phi}_\mathrm {tot,O} 
+ (1-\alpha) \Gamma_{max} ( 
{\bf{C}}_\mathrm {I},
{\bf{\Phi}}_\mathrm {tot,O}
)  \\
\end{split}
\label{eq:kaldi}
\end{equation}
and then distribute it to the between- and within-speaker covariance matrices.
Here, regularization is applied to the InD empirical total covariance matrix $\bf{C}_\mathrm {I}$ calculated from the unlabeled InD data and the total covariance of the OOD PLDA model by summing up the between- and within-speaker covariance 
\begin{equation}
\bf{\Phi}_\mathrm {tot,O} = \bf{\Phi}_\mathrm {B,O} + \bf{\Phi}_\mathrm{W,O}.
\label{eq:phi_tot}
\end{equation}
With regularization, the model guarantees that the total variances of $\bf {\Phi}_\mathrm{tot} $ will increase over that of $\bf {\Phi}_\mathrm{tot,O}$ before the adaptation.
After the adapted total covariance is obtained, we split it into the adapted between- and within-speaker covariance matrices:
\begin{equation}
\bf{\Phi}_\mathrm{W} ^{+}= \alpha\bf{\Phi}_\mathrm{W,O} 
+ \beta_\mathrm{W} \Gamma_{max} ( 
{\bf{C}}_\mathrm {I},
{\bf{\Phi}}_\mathrm {tot,O}
) \\
\label{eq:kaldi_w}
\end{equation}
\begin{equation}
\bf{\Phi}_\mathrm{B} ^{+}= \alpha \bf{\Phi}_\mathrm{B,O} + \beta_\mathrm{B} \Gamma_{max} ( 
{\bf{C}}_\mathrm {I},
{\bf{\Phi}}_\mathrm {tot,O}
).
\label{eq:kaldi_b}
\end{equation}
Here, $\beta_\mathrm{B} + \beta_\mathrm{W}  = (1-\alpha)$.
This adaptation was implemented as an unsupervised domain adaptation for Gaussian PLDA (G-PLDA) in the Kaldi toolbox \cite{kaldi}.  It is shown as special case 5 in Tab.~\ref{tab:generalform1}.
The procedure is illustrated in Algorithm~\ref{alg:kaldi}.


%

Both Kaldi's domain adaptation and CORAL+ are unsupervised domain adaptation methods that adapt OOD between- and within-speaker covariance matrices using unlabeled InD data. They utilize regularization to guarantee variance increases 
by choosing the larger value between two covariance matrices that represent the source domain, i.e., OOD, and the target domain, i.e., InD, respectively, in their simultaneous diagonalized subspace. They both use the empirical InD total covariance calculated from the InD data. 
The differences are: 
CORAL+ aims at variance increases in each of the between- and within-speaker covariances,
while Kaldi's aims at the variance increase only in the total covariance.
There is no guarantee of variance increases in both between- and within-speaker covariance. 
In regularization, Kaldi's uses empirical InD total covariance $\bf{C}_\mathrm{I}$ directly,
while CORAL+ uses pseudo InD within- and between-speaker covariance,
i.e., uses the empirical InD total covariance indirectly.
In Kaldi's method, after the total covariance has been adapted, two hyper-parameters $\{\beta_\mathrm{B},\beta_\mathrm{W}\}$ are set to determine the portions of the variance increases in the total covariance for the between- and within-speaker covariance.
The two hyper-parameters are restricted by each other and follow $\beta_\mathrm{B}+\beta_\mathrm{W}\le 1$.
CORAL+'s two hyper-parameters are independent from each other.

Regularization can also be applied to the supervised linear interpolation scenario.
Here, we extend the linear interpolations in \eqref{eq:lip} and \eqref{eq:lip_pseudo1} to 
\begin{equation}
\bf{\Phi^+}= \alpha {\bf{\Phi}}_{\mathrm{I}}  
+ (1-\alpha)\Gamma_\mathrm{max} ( {\bf{\Phi}}_{\mathrm{O}},{\bf{\Phi}}_{\mathrm{I}}   ),
\label{eq:lip_reg_IO}
\end{equation}
which is referred to as LIP with regularization (LIP reg) as proposed in~\cite{wang20} (special case 6 in Tab.~\ref{tab:generalform1}), and 
\begin{equation}
\bf{\Phi^+}= \alpha {\bf{\Phi}}_{\mathrm{I}}  
+ (1-\alpha)\Gamma_\mathrm{max} ( {\bf{\Phi}}_{\mathrm{I,pseudo}},{\bf{\Phi}}_{\mathrm{I}}   ),
\label{eq:lip_reg_I}
\end{equation}
respectively.
These two cases set the InD PLDA as the reference model for comparison purposes in order to confirm that applying regularization in the interpolation will ensure the uncertainty increase.
When we again use  CORAL \cite{sun16} to train the pseudo-InD PLDA as shown in \eqref{eq:coral_align3}, \eqref{eq:lip_reg_I} becomes the CIP with regularization (CIP reg) proposed in \cite{wang20}.  It is shown as special case 7 in Tab.~\ref{tab:generalform1}.

\subsubsection{Reintolarization in covariance alignment}
\label{ssset:reg1}

The regularization can also be applied to covariance alignment since it works with multiple covariance matrices as well. 
Rather than looking for an embedding-level transformation for the covariance alignment, as shown in \eqref{eq:coral_align1} in Section~\ref{ssec:correlation}, it seeks for a transformation so that the adapted OOD embeddings will have the covariance with a guaranteed uncertainty increase.
We can modify the embedding transformation in \eqref{eq:fda1} to 
\begin{equation}
\bm{\phi}_\mathrm{I, pseudo}  = 
\mathrm{\mathbf{C}_O^{\frac{1}{2}}
\mathbf{P}\hat{\Delta}^{\frac{1}{2}} \mathbf{P}{\trans} 
\mathbf{C}_O ^{-\frac{1}{2}}}
\bm{\phi}_\mathrm{O}
\label{eq:fda}
\end{equation}
by replacing the original $\Delta$  with a regularization $\hat{\Delta}= max (\mathbf{I}, \Delta)$.
Its two extremes are $\hat{\Delta}=\mathbf{I}$,
which corresponds to $\bm{\phi}_\mathrm{I,pseudo} =\bm{\phi}_\mathrm{O}$, meaning no adaptation is applied, 
and $\hat{\Delta}=\Delta$, where the embedding transformation is equivalent to \eqref{eq:fda1}. 
The algorithm is an unsupervised embedding-level domain adaptation, referred to as feature-distribution adaptor (FDA) in\cite{bousquet19}. It is shown as special case 8 in Tab.~\ref{tab:generalform1}.

In a way similar to Kaldi and CORAL+, FDA uses regularization to ensure uncertainty increase after the adaptation.
The difference is that it is an embedding-level transformation in rather than a model-level transformation like that seen in Kaldi and CORAL+. 
It has been argued that embedding-level domain adaptation is more relevant than model-level domain adaptation because the frequently used embedding pre-processing techniques, such as linear discriminant analysis (LDA) for discriminant dimensionality reduction \cite{bishop06} and within-class covariance normalization (WCCN), rely on the parameters learned by using adapted embeddings.

Since OOD data's labels are available, we can further replace the empirical total covariance $\mathbf{C}_\mathrm{O}$ in \eqref{eq:fda} with PLDA total covariance $\bf{\Phi}_\mathrm {tot,O}$ from \eqref{eq:phi_tot}
\begin{equation}
\begin{split}
\bm{\phi}_\mathrm{I, pseudo}  = &\mathbf{\Phi}_\mathrm{tot,O}^{\frac{1}{2}}\mathbf{P}\hat{\Delta}^{\frac{1}{2}} \mathbf{P}{\trans} \mathbf{\Phi}_\mathrm{tot,O} ^{-\frac{1}{2}}\bm{\phi}_\mathrm{O} \\
\mathbf{P}\hat{\Delta}\mathbf{P}{\trans}=&\Gamma_{max} ( \mathbf{\Phi}_\mathrm{tot,O} ^{-\frac{1}{2}} \mathbf{C}_\mathrm{I}  \mathbf{\Phi}_\mathrm{tot,O} ^{-\frac{1}{2}}, \mathbf{I}).
\end{split}
\label{eq:mk2}
\end{equation}
The corresponding modification of the PLDA between- and within speaker covariance matrices according to \eqref{eq:coral_align3}
are referred to as a modified Kaldi method (as special case 9 Kaldi* in Tab~\ref{tab:generalform1} and Sec~\ref{sec:exp}), which was also introduced in \cite{bousquet19} as well.
This method is an extension from FDA to model-level adaptation. 
When no pre-processing is used before model training, the difference in results arising from the application of the two methods will only be that arising from the changes in the use of OOD total covariance calculated from the OOD data or the trained OOD model. 

All of the above-mentioned methods are summarized as special cases in Tab.~\ref{tab:generalform1}. We show the relationships between them in Fig.~\ref{fig:relation} from the perspective of the three main factors of the generalized framework. For convenience, the relevant equations are shown in  Tab.~\ref{tab:eq}.

\label{ssec:coral}

\section{Experiments}
\label{sec:exp}
Experiments were conducted on the recent NIST SRE'16, 18, and 19 CTS datasets \cite{sre18}\cite{sre16}\cite{sre19}. 
Performance was evaluated in terms of equal error rate (EER) and minimum detection cost ($\rm minDCF$).

\subsection{Datasets}

The latest SREs organized by NIST have focused on domain mismatch as a particular technical challenge.
Switchboard, VoxCeleb 1 and 2, and MIXER corpora that consisted of SREs 04 – 06, 08, 10, and 12 were used to train an x-vector extractor. 
They were considered to be OOD data as they are English speech corpora,
while SRE'16 is in Tagalog and Cantonese, and SRE'18 and 19 are in Tunisian Arabic.

Data augmentation was   applied to the training set in the following ways:
(a) Additive noise: each segment was mixed with one of the noise samples in the PRISM corpus \cite{ferrer11} (SNR: 8, 15, or 20dB), 
(b) Reverberation: each segment was convolved with one of the room impulse responses in the REVERB challenge data \cite{kinoshita16}, and
(c) Speech encoding: each segment was encoded with AMR codec (6.7 or 4.75 kbps)\cite{lee19a}.

Only the MIXER corpora and its augmentation, which consisted of 262,427 segments from 4,322 speakers in total, were used as OOD data in PLDA training.
The development data and evaluation data used are shown in Tab.~\ref{tab:data}.
For SRE'16, only the \textit{cmntrain unlabeled major} set (2,272 segments) was available as a development set.
Thus, only unsupervised domain adaptation experiments were conducted on SRE'16. 
For the two partitions Tagalog and Cantonese in SRE'16, only equalized results are shown in this paper.
SRE'18 has three datasets:  an evaluation set (188 speakers, 13,451 segments), 
a development set (25 speakers, 1,741 segments),
and an unlabeled set (2,332 segments). 
For SRE'19 there was no development set available, but it was collected together with SRE'18. Thus, they are considered to be in the same domain. Therefore, we followed the common benchmark setting and use the SRE'18 evaluation set as the development data for SRE'19 experiments. To use the same setting for SRE'18 experiments, we evaluated the development set.
In this section, we refer to the evaluation dataset as InD data, and to the development set as enroll and test data, in order to avoid confusion. In the following experiments, we used SRE16, SRE18, and SRE19 to specifically present the experiments using the above-mentioned settings to differ from the datasets SRE'16, SRE'18 and SRE'19.

\begin{table}[t]
\caption{Development data and evaluation data used in the experiments }
\begin{center}
\begin{tabular}{c|  c | c| c}
\hline
Experiments & OOD data & InD data  &  Eval data \\
\hline
SRE16 & MIXER & cmntrain unlabeled major  & cmneval \\
SRE18 & MIXER & cmn2eval & cmn2dev \\
SRE19 & MIXER & cmn2dev  & SRE'19 eval \\
\hline
\end{tabular}
\label{tab:data}
\end{center}
\end{table}

\subsection{Experimental setup}

To confirm the generalization of the proposed methods, we used two Time-delayed neural network structures for the x-vector extraction: a 43-layer TDNN (deep) and a 27-layer TDNN (shallow). 
Both have residual connections and 
a 2-head attentive statistics pooling, in the same way as in \cite{lee19a}.
Additive margin softmax loss (s=40.0, m=0.15) was used for the optimization \cite{wang18}.
The number of dimensions of the x-vector was 512. 
The mean shift was applied to OOD data using its mean. 
InD data and enroll and test data were centralized using InD data.
LDA was used in some of the G-PLDA systems to reduce dimensionality to 150-dimension. The usage is clarified with the experiments later. 
In our interpolation domain adaptation experiments where LDA was applied, the LDA projection matrix was calculated from the training data for the base PLDA, $\mathbf{\Phi_0}$ in (\ref{eq:general}), and applied to the training data for all the PLDA in the interpolation, as well as the evaluation data.
For the single InD G-PLDA and OOD G-PLDA training, LDA matrices were calculated from the InD data and OOD data, respectively. LDA is not used in any HT-PLDA systems. Our implementation of domain adaptation for HT-PLDA is based on \cite{anna18}, and $\lambda$ is set as 2 as the default setting in the code\footnote{\url{https://github.com/bsxfan/meta-embeddings/tree/master/code/Niko/matlab/clean/VB4HTPLDA}}.

We report here results in terms of equal error rate (EER)
and the minimum of the normalized detection cost function (minDCF),
for which we assume two prior target probabilities $P_\mathrm{Target}$ of 0.01 and 0.005, and equal weights of 1.0 between misses $C_\mathrm{Miss}$ and false alarms $C_\mathrm{FalseAlarm}$,
as defined in the evaluation plan of NIST SRE' 16, 18, and 19 \cite{sre18}\cite{sre16}\cite{sre19}. 
In the later experiments, we use, rather, $\rm min\rm C_{primary}$, which is averaged from the two minDCFs.


\subsection{G-PLDA vs. HT-PLDA}
\label{ssec:results}



\begin{table}
\caption{Performance of G-PLDA and HT-PLDA on SRE'16-19. LDA was not applied.}
\label{tab:exp0}
\centering

\begin{subtable}{0.5\textwidth}
\centering
\caption{EER(\%)}
   \label{tab:exp0a}
\begin{tabular}[t]{l|cc|cc}
\hline
     & \multicolumn{2}{c|}{Shallow} & \multicolumn{2}{c}{Deep} \\
     & G-PLDA & HT-PLDA & G-PLDA & HT-PLDA  \\ 
\hline \hline
SRE16 OOD               & 8.17  & 7.98 & 5.78 & 5.49 \\
SRE18 OOD               & 8.02  & 7.21 & 6.37 & 5.15\\
SRE18 InD               & 7.49  & 5.59 & 4.39 & 3.93 \\
SRE19 OOD               & 7.47  & 6.95 & 4.58 & 3.86 \\
SRE19 InD               & 6.20  & 5.29 & 4.00 & 3.30   \\
\hline
\end{tabular}
\end{subtable}

\bigskip
\begin{subtable}{0.5\textwidth}
\centering
\caption{$\mathrm{minDCF}_1$/$\mathrm{minDCF}_2$ using $P_\mathrm{Target}$ equivalent to 0.01 and 0.005, respectively} 
   \label{tab:exp0b}
   
\begin{tabular}[t]{l|cc|cc}
\hline
     & \multicolumn{2}{c|}{Shallow} & \multicolumn{2}{c}{Deep} \\
     & G-PLDA & HT-PLDA & G-PLDA & HT-PLDA  \\ 
\hline \hline
SRE16 OOD              & 0.662/0.762 & 0.560/0.621 & 0.462/0.539 & 0.433/0.491\\
SRE18 OOD              & 0.518/0.556 & 0.456/0.512 & 0.411/0.440& 0.382/0.424\\
SRE18 InD              & 0.400/0.447 & 0.337/0.376 & 0.256/0.298 & 0.224/0.262\\
SRE19 OOD              & 0.538/0.601 & 0.513/0.577 & 0.364/0.424 & 0.329/0.395\\
SRE19 InD              & 0.501/0.567 & 0.420/0.477 & 0.313/0.369 & 0.258/0.302 \\
\hline
\end{tabular}
\end{subtable}
\end{table}
\begin{table}[t]
\caption{Investigations of an LDA application to G-PLDA systems. The projection matrix is computed from raw x-vectors or adapted features; ``*-1'' indicates those whose LDA were calculated from the adapted features, while ``*-2'' represents those whose LDA were calculated from the raw x-vectors. X-vectors were extracted from the 43-layer network.} Results are shown as EER/$\rm min$$\rm C_{primary}$.
\centering
\begin{tabular}[t]{l|c|c|c}
\hline
     & SRE16 & SRE18 & SRE19 \\
\hline \hline
InD (no LDA)            & - / -               & 4.39/0.277          & 4.00/0.341 \\
OOD (no LDA)            & 5.78/0.500          & 6.37/0.426         & 4.58/0.394  \\
\hline
InD                     & - / -                 & 4.52/0.278          & 4.04/0.361 \\
OOD                     & 5.84/0.494           & 6.18/0.415           & 4.53/0.394    \\
\hline
CORAL-1                  & 5.67/0.410           & 5.15/0.279           & 4.90/0.390    \\
CORAL-2                  & \textbf{4.03/0.353}  & \textbf{4.31/0.251}  & \textbf{3.89/0.316}    \\
\hline
FDA-1                    & 4.17/\textbf{0.331}  & 4.35/\textbf{0.228}  & 3.54/\textbf{0.288}    \\
FDA-2                    & \textbf{3.76}/0.335  & \textbf{4.22}/0.244  & \textbf{3.50}/0.298   \\
\hline
\end{tabular}
\label{tab:exp_lda}
\end{table}

We first evaluated the single G-PLDA and HT-PLDA trained from the mismatched (OOD) and matched (InD) data, using the two x-vector extractor structures. LDA was not applied. Tables~\ref{tab:exp0a} and \ref{tab:exp0b} show performance using x-vectors extracted from the 27-layer TDNN (shallow) and 43-layer TDNN (deep), respectively. 
The two minDCF values were calculated using the two operating points in which $P_\mathrm{Target}$ equals to 0.01 and 0.005, respectively. They represent two points in the DET curve.
A comparison of performance using HT-PLDA and G-PLDA within each of Tab.~\ref{tab:exp0a} and \ref{tab:exp0b} shows that HT-PLDA outperformed G-PLDA by a large margin in EER and both minDCFs, 
for both OOD and InD domains.
This effect remains the same when using either of the front-ends.
Such observations are consistent with prior studies.
This, again, demonstrates that it is essential to develop domain adaptation for HT-PLDA, which is the better baseline. 

A comparison of the two tables shows that using a deeper neural network yields greater improvement in performance in G-PLDA systems then that in HT-PLDA systems. 
The average EER/$\rm min\rm C_{primary}$ reduction in G-PLDA systems is $33.0\%/29.4\%$, 
while it is $29.1\%/25.5\%$  in HT-PLDA systems.

The effectiveness of HT-PLDA can be seen to be larger in shallower networks than in deeper networks:
$17.11\%/15.2\%$ and $13.6\%/11.5\%$ average reduction in EER/$\rm min\rm C_{primary}$ using the shallower network and deep network, respectively.
Since the two minDCFs show consistent results, in the later experiments we instead used only $\rm min\rm C_{primary}$, which is averaged from the two. 
Since the systems with the deeper TDNN outperform those with the shallow one in all settings, we will next focus on and show only the experiments using the 43-layer TDNN (deep). The corresponding results for the 27-layer TDNN (shallow) are shown in Appendix B.

\begin{table}[t]
\caption{Domain adaptations in SRE16-SRE19 with two TDNN structured x-vector extractors. X-vectors were extracted from the 43-layer network. Results are shown as EER/$\rm min$$\rm C_{primary}$.}
\centering

\begin{subtable}{0.5\textwidth}
\centering
\caption{SRE16}
   \label{tab:exp1_sre16}
   \begin{tabular}[t]{l|c|c}
\hline
     & G-PLDA & HT-PLDA \\
\hline \hline
OOD                     & 5.84/0.494  & 5.49/0.462 \\
\hline
CORAL                   & 4.03/0.353  & 5.64/0.470\\
FDA                     & \textbf{3.76}/\textbf{0.335}  
                        & \textbf{4.66}/0.426\\
CORAL+                  & 4.01/0.350  & 4.77/0.449\\
Kaldi                   & 4.04/0.342  & 4.96/\textbf{0.410}\\
Kaldi*                  & 3.92/0.336  & 4.69/0.428\\
\hline
\end{tabular}
\end{subtable}

\bigskip
\begin{subtable}{0.5\textwidth}
\centering
\caption{SRE18 and SRE19. The weights in all the interpolations were chosen to be 0.5.} 
\label{tab:exp1_sre19}
\begin{tabular}[t]{l|cc|cc}
\hline
\multirow{2}{4em}{System}     & \multicolumn{2}{c|}{SRE18} & \multicolumn{2}{c}{SRE19} \\
     & G-PLDA & HT-PLDA & G-PLDA & HT-PLDA \\ 
\hline \hline
OOD                     & 6.18/0.415 & 5.15/0.403 & 4.53/0.394   & 3.86/0.362   \\
InD                     & 4.52/0.278 & 3.93/0.243 & 4.04/0.361    & 3.30/0.280   \\
\hline
CORAL                   & 4.31/0.251 & 4.92/0.334 & 3.89/0.316     & 3.95/0.346   \\
FDA                     & 4.22/0.244 & 4.11/0.281 & 3.50/0.298     & 3.23/0.297   \\
CORAL+                  & 4.31/0.264 & 4.14/0.282& 3.34/0.305     & 3.23/0.307   \\
Kaldi \cite{kaldi}      & 4.20/0.271 & 4.23/0.280 & 3.82/0.315     & 3.77/0.350   \\
Kaldi*                  & 4.20/0.251 & 4.41/0.345  & 3.40/0.300     & 3.19/0.295   \\
\hline
LIP(OOD)                & 3.85/0.222  & 3.41/0.222 & 3.12/0.286  &  2.56/0.239   \\
LIP(CORAL)              & 3.89/0.189  & 3.36/0.198 & 3.19/0.291  &   2.47/\textbf{0.222}  \\
LIP(FDA)                & 3.80/0.200  & \textbf{3.29}/0.213 & 3.02/0.281  &    2.55/0.230 \\
LIP(CORAL+)             & 3.68/0.212  & 3.41/0.201& 2.90/0.275  &   2.46/0.228   \\
LIP(Kaldi)              & 3.83/0.205  & 3.38/0.241  & 3.03/0.276  &    2.57/0.249  \\
LIP(Kaldi*)             &  \textbf{3.64}/0.213  & 3.31/0.210& 2.93/0.273  &    \textbf{2.42}/0.225 \\
\hline
LIP-reg(OOD)            & 3.86/\textbf{0.195}  & 3.63/0.202& 3.33/0.291  &    2.66/0.237   \\
LIP-reg(CORAL)          & 3.77/0.199  & 3.38/\textbf{0.189}  & 3.08/0.279  &    2.55/0.223  \\
LIP-reg(FDA)            & 4.36/0.210 & 3.43/0.201 & \textbf{2.88}/\textbf{0.269}   & 2.42/0.225 \\
LIP-reg(CORAL+)         & 3.78/\textbf{0.195} & 3.58/0.194 & 3.08/0.283  &   2.58/0.232   \\
LIP-reg(Kaldi)          & 3.73/0.220  & 3.36/0.230  & 3.13/0.280  &   2.61/0.244\\
LIP-reg(Kaldi*)         & 3.78/0.202  & 3.43/0.199 & 3.01/0.279  &  2.55/0.230\\
\hline
\end{tabular}
   
\end{subtable}
 \label{tab:exp1}
\end{table}

\subsection{G-PLDA and HT-PLDA with domain adaptations}
\begin{figure}[t]
\centering
\includegraphics[width=0.9\linewidth]{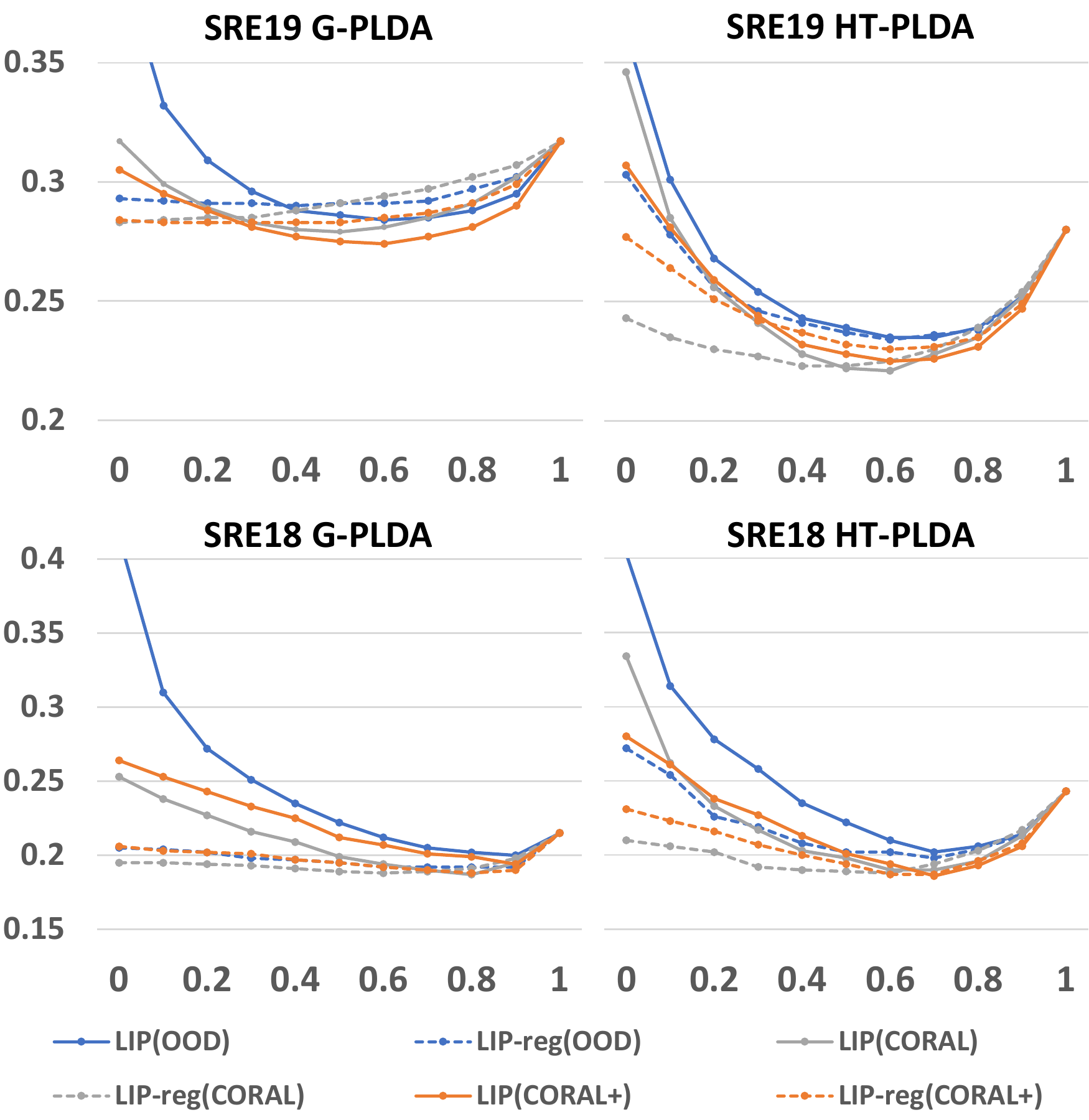}
\caption{ 
$\rm min$$\rm C_{primary}$ of SRE18 and SRE19 evaluations using G-PLDA and HT-PLDA with varying interpolation weights. Three interpolation settings without and with regularizations are compared: (1) LIP between OOD and InD PLDA, (2) LIP between CORAL adapted PLDA and InD PLDA, and (3) LIP between CORAL+ adapted PLDA and InD PLDA. The regularizations use the InD PLDA as a reference. The x-vectors were extracted from the 43-layer network. X-axis shows the interpolation weight for InD PLDA, and the y-axis shows $\rm min$$\rm C_{primary}$. }
  \label{fig:reg-gplda}
\end{figure}
We evaluated the two PLDAs with embedding-level adaptation methods, i.e., CORAL+ and FDA,  as noted in Section~\ref{sec:sec4}.
In G-PLDA, LDA and length normalization were applied to the x-vectors as pre-processing 
before PLDA evaluations.
Thus, an LDA projection matrix can be calculated from either the raw x-vectors from which the CORAL transformation and FDA transformation were also derived, or from the transformed x-vectors after CORAL and FDA have been applied.
We investigated the use of both LDA matrices (see  Tab.~\ref{tab:exp_lda}).
It is worth noticing that LDA was applied to the OOD and InD PLDA systems here. Therefore, their performance differs from that in Tab.~\ref{tab:exp0} which is referred to as InD (no LDA) and OOD (no LDA) in Tab.~\ref{tab:exp_lda}.
Notably, in CORAL, we found that  ``CORAL-2'', which used LDA calculated from the raw x-vectors, achieved the best performance in both measurements in the three data sets.
For FDA, the advantage was shown only in EER, while minDCFs were slightly worse.
For the later experiments of embedding-level domain adaptation for G-PLDA, we used LDA calculated from the raw x-vectors. 

We show in Tab.~\ref{tab:exp1} several special cases of the generalized format of domain adaptation for the three datasets.
Only unsupervised methods are shown for SRE16 evaluations.
It first shows OOD PLDAs' performance results, for reference purposes, and five existing unsupervised domain adaptation methods' results, including those for CORAL, FDA, CORAL+, Kaldi adaptation, and the modified Kaldi. 
All five adaptation methods employed the correlation alignment factor of the generalized framework, and the latter four also employ the covariance regularization factor. The CORAL adaptation achieved improvements over those of the OOD results in the case of G-PLDA systems for all three datasets, while failing in working with HT-PLDA.  The rest four adaptation methods significantly improved both G-PLDA and HT-PLDA systems for all three datasets. The results indicate the effectiveness of the covariance regularization factor, especially for HT-PLDAs. 
For SRE18 and SRE19 evaluations where the development data is available to train InD PLDA models, Tab.~\ref{tab:exp1_sre19} shows that the G-PLDA with the above-mentioned unsupervised adaptation methods even outperformed InD G-PLDA system. 
For HT-PLDA,  the application of FDA, CORAL+, and Kaldi* also outperformed the InD HT-PLDA in terms of EER for SRE19 evaluation, while a slight increase in minDCF. For the SRE18 evaluation, the application of the unsupervised adaptation methods did not bring improvement.
To be noted, LDA was applied to all the G-PLDA systems in Tab.~\ref{tab:exp1}. Therefore, their baseline OOD and InD G-PLDA results are consistent with those in Tab.~\ref{tab:exp_lda} rather than those in Tab.~\ref{tab:exp0}.

We next investigated the special cases specifically using the covariance matrix interpolation (LIP) factor of the generalized framework.
Since InD PLDA outperformed OOD PLDA for both types of PLDA in the evaluations of both SRE18 and SRE19, we used it as the base PLDA and interpolated it with the above-mentioned unsupervised-adapted PLDA. 
As shown in Tab.~\ref{tab:exp1}, LIP and LIP-reg systems interpolate InD PLDA and the unsupervised-adapted PLDAs shown in the brackets in the “System” column. For LIP-reg systems, the InD PLDA is used as the reference PLDA for the regularization of the unsupervised-adapted PLDAs shown in the brackets before being interpolated.
The weights in all the interpolations in Tab.~\ref{tab:exp1} were chosen to be $0.5$. 
LDA projection matrix was calculated from the InD data and applied to all the LIP systems.
We observed that all the LIP systems outperformed their single systems, and the observation is consistent for both G-PLDA and HT-PLDA and for both datasets. We conclude that the adaption by the linear interpolation with the InD PLDA further improved the performance over the unsupervised domain-adapted PLDAs.

On the basis of the LIP systems, we show their corresponding systems (LIP-reg) with a covariance regularization factor of the generalized framework. The InD PLDA was additionally used as the reference PLDA, $\mathbf{\Phi}_2$ in (\ref{eq:general}), for the regularization purpose.  We observed that for both PLDAs, using regularization achieved similar performance to that without regularization. Above is the analysis using 0.5 as the interpolation weight, and next, we looked into different weights.

\subsection{Robustness of domain adaptations using regularization}
To further confirm the effectiveness of the regularization factor in the linear fusions on speaker verification performance, we further investigated the robustness of the performance against varying interpolation weights. 
We take the next three settings as examples: LIP between OOD and InD PLDA, LIP between CORAL adapted PLDA and InD PLDA,  LIP between CORAL+ adapted PLDA and InD PLDA, and these three settings with regularizations using InD PLDA as a reference. 
The plots in Fig.~\ref{fig:reg-gplda} show that although in the same settings the optimal $\rm min\rm C_{primary}$ may be better when not using regularization, those with regularization generally were less sensitive to the interpolation weights. We observe in all the four plots that a bigger range of interpolation weights achieved lower $\rm min\rm C_{primary}$  when regularization was applied. 
Specifically, with the varying interpolation weights from 0 to 1, the minDCFs have the range 0.223--0.333 on average over all the LIP systems in Fig. 4, while the range for LIP-reg systems is smaller, i.e., 0.224--0.268, with a lower maximum minDCF as 0.268.  The average minDCF of the LIP-reg systems is 0.237, lower than that of the LIP systems at 0.253. The standard deviation of the LIP-reg systems is 0.013, much smaller than that of the LIP systems at 0.032. 
Thus, we conclude that the regularization improved robustness for interpolations w.r.t. varying interpolation weights. This was constant in both G-PLDA and HT-PLDA experiments.

\section{conclusion}
\label{sec:summary}

We have proposed here a generalized framework for domain adaptation of PLDA in speaker recognition that works with both unsupervised and supervised methods, 
as well as two new techniques: (1) correlation-alignment-based interpolation and (2) covariance regularization.
The generalized framework enables us to combine the two techniques and also several existing supervised and unsupervised domain adaptation methods into a single formulation. 
Further, our proposed regularization technique ensures robustness for interpolations w.r.t. varying interpolation weights, which in practice is essential. 
In future, we intend to evaluate this approach's effectiveness on other DNN-based speaker embeddings.

\ifCLASSOPTIONcaptionsoff
  \newpage
\fi



\bibliographystyle{IEEEtran}
\bibliography{refs}
%



%




\clearpage
\appendices
\section{CORAL+ algorithm}
\label{sec:apxA}
CORAL+\cite{lee19b} (case 4 in Tab.~\ref{tab:generalform1}) uses CORAL~\cite{sun16} to train the pseudo-InD PLDA as shown in \eqref{eq:coral_align3} and interpolate with the OOD PLDA covariance matrix with regularization. The procedure is illustrated next:
\begin{algorithm}[h]
\SetAlgoLined
  \textbf{Input} 
    Out-of-domain PLDA matrices $\{\bf{\Phi}_\mathrm{B,O},\bf{\Phi}_\mathrm{W,O}\}$\\
    \hspace*{\algorithmicindent} In-domain data $X_{I}$\\
    \hspace*{\algorithmicindent}Adaptation hyper-parameters $\{\gamma,\beta\}$ \\
  \textbf{Output} 
    Adapted covariance matrices $\{\bf{\Phi}_\mathrm{B},\bf{\Phi}_\mathrm{W}\}$\\

Estimate empirical covariance matrix from the in-domain data $X_{I}$\\
\hspace*{\algorithmicindent}    $\mathbf{C}_\mathrm{I}=\textsc{Cov}(X_{I})$\\
Compute out-of-domain covariance matrix\\
\hspace*{\algorithmicindent}   $\bf {C}_\mathrm {O}=\bf {\Phi}_\mathrm {B,O} +\bf{\Phi}_\mathrm {W,O}$ \\
\textbf{for each $\bf{\Phi}$ in $\{\bf{\Phi}_\mathrm{B,O}, \bf{\Phi}_\mathrm{W,O}\}$ do}
\\
\hspace*{\algorithmicindent}Compute Pseudo in-domain covariance matrix\\
\hspace*{\algorithmicindent} 
$\bf{S=C}_\mathrm{I}^{1/2} \bf{ C}_\mathrm{O}^{-1/2}\Phi C_\mathrm{O}^{-1/2} C_I^{1/2}$\\
\hspace*{\algorithmicindent}
Find $\bf{\{B,E\}}$ via simultaneous diagonalization of $\bf{\Phi}$ and $\bf{S}$ \\
\hspace*{\algorithmicindent}$
\{\bf{Q,\Lambda\}} \leftarrow EVD (\bf{\Phi})$ \\
\hspace*{\algorithmicindent}$
\{\bf{P,E\}} \leftarrow EVD (\bf{\Lambda} ^ \mathrm{-1/2} Q{\trans} S Q \Lambda ^ \mathrm{-1/2})$ \\
\hspace*{\algorithmicindent}$\bf{B=Q\Lambda} ^ \mathrm{-1/2} P$\\
CORAL+ unsupervised adaptation of PLDA, $\alpha \in \{\gamma,\beta\}$ \\
\hspace*{\algorithmicindent}$\bf{\Phi} = \bf{\Phi}_\mathrm{O} + \gamma \bf{B}^{\mathsf{-T}}\max\bf{(E-I)B}^\mathrm{-1} $\\

\textbf{Notation} $\mathrm{EVD}(\cdot)$ returns a matrix of eigenvectors and the corresponding eigenvalues in a diagonal matrix.
 \caption{The CORAL+ algorithm for unsupervised adaptation of PLDA.}
\label{alg:coralp}
\end{algorithm}
\vspace*{-3mm}
\section{Experimental results with 27-layer TDNN}
\label{sec:exp27}

We show the experimental results using the x-vectors extracted from the 27-layer TDNN (shallow). The corresponding results for the 43-layer TDNN (deep) are shown in Tab.~\ref{tab:exp_lda} and \ref{tab:exp1} and Fig.~\ref{fig:reg-gplda}.

Table~\ref{tab:exp_lda_27} investigates the effect of LDA in G-PLDA (corresponding to Tab.~\ref{tab:exp_lda}). The LDA projection matrix was computed from raw x-vectors or adapted embedings; ``*-1'' and ``*-2'' indicate that LDA were calculated from the adapted embeddings and the raw x-vectors, respectively.
\begin{table}[h]
\caption{Results are shown as EER/$\rm min$$\rm C_{primary}$.}
\centering
\begin{tabular}{l|c|c|c}
\hline
     & SRE16 & SRE18 & SRE19 \\
\hline \hline
InD                      & - / -                 & 7.53/0.435          & 6.33/0.561\\
\hline
OOD                      & 8.15/0.701           & 7.57/0.526           & 7.29/0.562    \\
\hline
CORAL-1                  & 6.85/0.507           & 6.49/0.380           & 6.67/0.553    \\
CORAL-2                  & \textbf{5.62/0.453}  & \textbf{5.45/0.366}  & \textbf{5.77/0.494}    \\
\hline
FDA-1                    & 5.54/\textbf{0.433}  & 5.34/\textbf{0.339}  & 5.69/0.438   \\
FDA-2                    & \textbf{5.36}/0.443  & \textbf{5.13}/0.352  & \textbf{5.68}/\textbf{0.410}    \\
\hline
\end{tabular}
\label{tab:exp_lda_27}
\end{table}

Table~\ref{tab:exp1_27} shows the domain adaptations using the special cases in SRE16-SRE19 (corresponding to Tab.~ \ref{tab:exp1}). For G-PLDA, we used LDA calculated from the raw x-vectors.

\begin{figure}[t]
\centering
\includegraphics[width=0.9\linewidth]{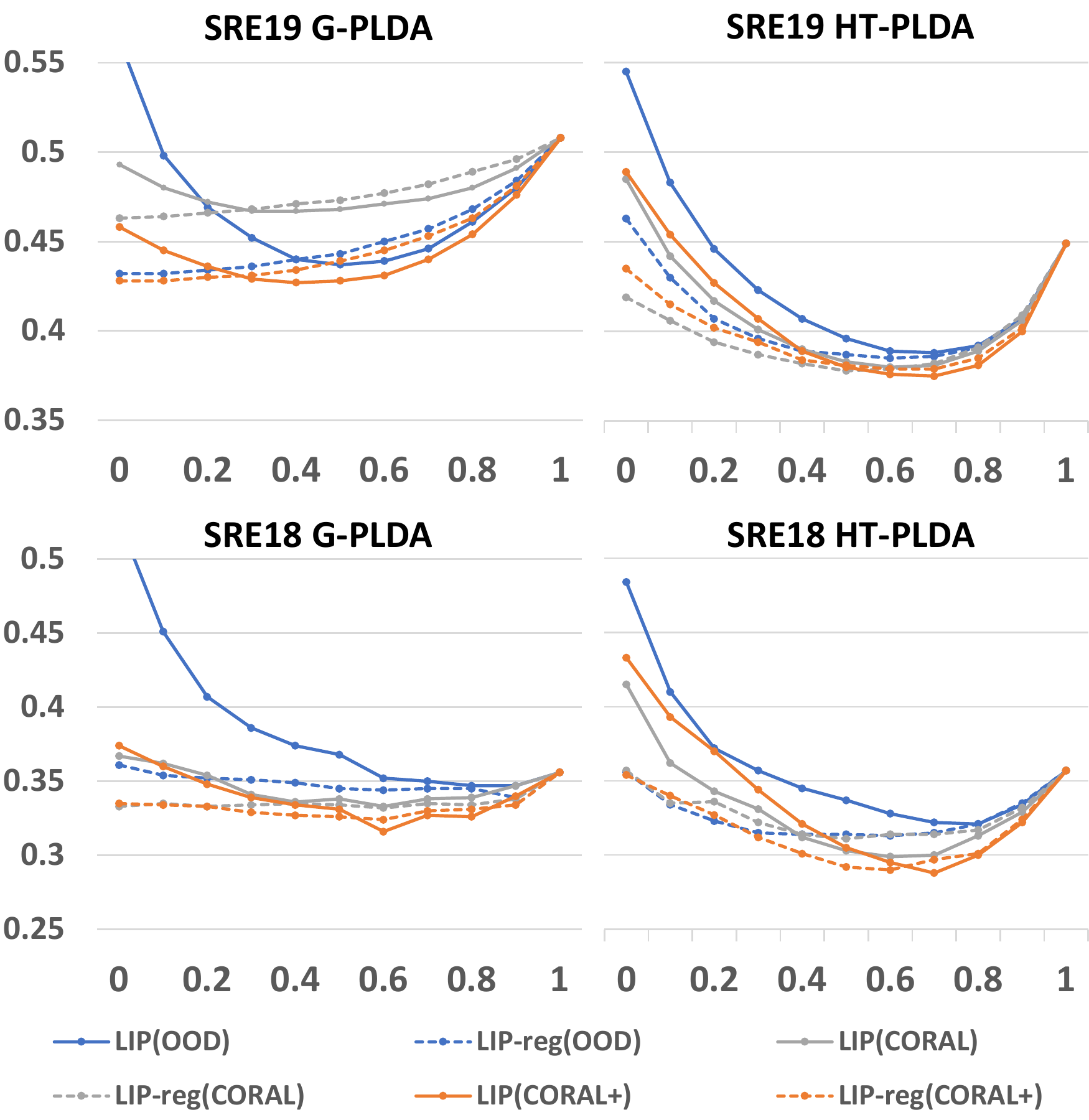}
\caption{ $\rm min$$\rm C_{primary}$ of SRE18 and SRE19 evaluations using G-PLDA and HT-PLDA.}
  \label{fig:plot-27}
\end{figure}

Figure~\ref{fig:plot-27} investigates the robustness of the performance against varying interpolation weights for shallow x-vectors (corresponding to Fig.~\ref{fig:reg-gplda}).
\begin{table}[h]
\caption{Domain adaptations in SRE16-SRE19. Results are shown as EER/$\rm min$$\rm C_{primary}.$}
\centering
\begin{subtable}{0.5\textwidth}
\centering
\caption{SRE16}
   \label{tab:exp2_sre16}
   \begin{tabular}[h]{l|cc}
\hline
    
     & G-PLDA & HT-PLDA \\ 
\hline \hline
OOD                     & 8.15/0.701   & 7.98/0.591 \\
\hline
CORAL                   & 5.62/0.453   & 8.11/0.569 \\
FDA                     & \textbf{5.36}/\textbf{0.443}   & 7.27/0.548 \\
CORAL+                  & 5.60/0.464  & 7.65/0.580 \\
Kaldi     & 5.55/0.452  & \textbf{7.18}/\textbf{0.499} \\
Kaldi*                  & 5.48/0.445 & 7.26/0.548 \\
\hline
\end{tabular}
\end{subtable}

\bigskip
\begin{subtable}{0.5\textwidth}
\centering
\caption{SRE18 and SRE19}\label{tab:exp1_sre19_27}
\begin{tabular}[t]{l|cc|cc}
\hline
     & \multicolumn{2}{c|}{SRE18} & \multicolumn{2}{c}{SRE19} \\
     & G-PLDA & HT-PLDA & G-PLDA & HT-PLDA \\ 
\hline \hline
OOD                     & 7.57/0.526 & 7.21/0.484  & 7.29/0.562 & 6.95/0.545\\
InD                     & 7.53/0.435 & 5.59/0.356  & 6.33/0.561 & 5.29/0.449\\
\hline
CORAL                   & 5.45/0.366 & 6.40/0.415  & 5.77/0.494 & 6.12/0.485\\
FDA                     & 5.13/0.352 & 6.13/0.400  & 5.68/0.410 & 5.88/0.476\\
CORAL+                  & 5.41/0.374 & 6.29/0.433  & 5.53/0.458 & 5.99/0.489\\
Kaldi \cite{kaldi}      & 5.35/0.361 & 7.36/0.447  & 5.88/0.455 & 7.26/0.548\\
Kaldi*                  & 5.26/0.364 & 6.16/0.400  & 5.62/0.449 & 5.90/0.477\\
\hline
LIP(OOD)                & 5.19/0.369 & 4.76/0.337 & 5.05/0.437 & 4.56/0.396\\
LIP(CORAL)              & 5.09/0.334 & 4.75/0.303 & 5.19/0.473 & 4.55/0.383\\
LIP(FDA)                & 4.83/0.315 & 4.58/0.312 & 4.92/0.447 & 4.36/0.383\\
LIP(CORAL+)             & 4.76/0.331 & 4.57/0.305 & 4.77/0.428 & 4.53/0.380\\
LIP(Kaldi)              & 4.76/0.318 & 4.92/0.337 & 4.88/0.431 & 4.65/0.423\\
LIP(Kaldi*)             & \textbf{4.67}/\textbf{0.313} & 4.53/0.311 & 4.77/0.426 & 4.37/0.383\\
\hline
LIP-reg(OOD)            & 5.30/0.346 & 4.57/0.314 & 5.26/0.444 & 4.57/0.387\\
LIP-reg(CORAL)          & 4.84/0.338 & 4.78/0.311 & 4.98/0.468 & 4.69/\textbf{0.378}\\
LIP-reg(FDA)            & 4.68/0.315 & 4.58/0.304 & \textbf{4.72}/\textbf{0.425} &4.52/0.383 \\
LIP-reg(CORAL+)         & 4.91/0.326 & \textbf{4.48}/\textbf{0.292} & 4.97/0.439 & \textbf{4.13}/0.403\\
LIP-reg(Kaldi)          & 4.92/0.316 & 4.82/0.320 & 4.97/0.440 & 4.53/0.408\\
LIP-reg(Kaldi*)         & 4.81/0.317 & 4.58/0.302 & 4.89/0.437 &4.51/0.382\\
\hline
\end{tabular}
\end{subtable}
 \label{tab:exp1_27}
\end{table}
\vspace{-3mm}

\clearpage

\end{document}